\documentclass[
  journal=jpccck,
  manuscript=article,
  layout=twocolumn]{achemso}

\usepackage[fontsize=10pt]{scrextend}
\usepackage[T1]{fontenc}
\usepackage[utf8]{inputenc}

\usepackage{xr}
\usepackage[separate-uncertainty=true]{siunitx}
\usepackage{amsmath}
\usepackage{mathtools}
\usepackage{balance}
\usepackage[colorlinks=true,
    citecolor=blue,
    linkcolor=blue,
    urlcolor=blue,pagebackref=false]{hyperref}
\usepackage[version=3]{mhchem}
\usepackage[pagewise]{lineno}

\setkeys{acs}{doi = true}
\newcommand{\doi}[1]{\href{http://dx.doi.org/#1}{\nolinkurl{#1}}}

\graphicspath{
  {Figs/}
  {../_JAM_MS__Tirmzi201708__Perovskite_KS_dissipation__figs/}
  {../_JAM_MS__Tirmzi201708__Perovskite_KS_dissipation__figs/Figs/}
  {/Users/ryandwyer/Dropbox/_JAM_MS__Tirmzi201708__Perovskite_KS_dissipation__figs/}
  {/Users/ryandwyer/Dropbox/_JAM_MS__Tirmzi201708__Perovskite_KS_dissipation__figs/Figs/}
}

\newcommand{\st}[1]{_{\mathrm{#1}}} 
\SectionNumbersOn
\DeclareSIUnit{\molar}{M}
\DeclareSIUnit{\sq}{sq.}
\renewcommand{\Re}{\operatorname{Re}}

\DeclarePairedDelimiter\abs{\lvert}{\rvert}

        \title{Substrate-Dependent Photoconductivity Dynamics in a High-Efficiency Hybrid Perovskite Alloy}
    \author{Ali Moeed Tirmzi}
        \affiliation{Dept.\ of Chemistry and Chemical Biology, Cornell University, Ithaca, NY 14853, USA}
        \author{Jeffrey A. Christians}
        \affiliation{National Renewable Energy Laboratory, Golden, CO 80401, USA}
        \author{Ryan P. Dwyer}
        \affiliation{Dept.\ of Chemistry and Chemical Biology, Cornell University, Ithaca, NY 14853, USA}
        \author{David T. Moore}
        \affiliation{National Renewable Energy Laboratory, Golden, CO 80401, USA}
        \author{John A. Marohn}
        \affiliation{Dept.\ of Chemistry and Chemical Biology, Cornell University, Ithaca, NY 14853, USA}
        \email{jam99@cornell.edu}

\begin{document}

\nolinenumbers 
\sloppy 


\begin{abstract}

Films of \ce{(FA_{0.79}MA_{0.16}Cs_{0.05})_{0.97}Pb(I_{0.84}Br_{0.16})_{2.97}} were grown over \ce{TiO2}, \ce{SnO2}, \ce{ITO}, and \ce{NiO}.
Film conductivity was interrogated by measuring the in-phase and out-of-phase forces acting between the film and a charged microcantilever.
We followed the films' conductivity \latin{vs}.\ time, frequency, light intensity, and temperature (\SI{233}{} to \SI{312}{\kelvin}).
Perovskite conductivity was high and light-independent over \ce{ITO} and \ce{NiO}.
Over \ce{TiO2} and \ce{SnO2}, the conductivity was low in the dark, increased with light intensity, and persisted for 10's of seconds after the light was removed.
At elevated temperature over \ce{TiO2}, the rate of conductivity recovery in the dark showed an activated temperature dependence ($E\st{a}  = \SI{0.58}{\text{eV}}$).
Surprisingly, the light-induced conductivity over \ce{TiO2} and \ce{SnO2} relaxed essentially instantaneously at low temperature.
We use a transmission-line model for mixed ionic-electronic conductors to show that the measurements presented are sensitive to the sum of electronic and ionic conductivities. 
We rationalize the seemingly incongruous observations using the idea that holes, introduced either by equilibration with the substrate or \latin{via} optical irradiation, create iodide vacancies. 

\end{abstract}



\section{Introduction}

The extraordinary performance of solar cells made from solution-processed lead-halide perovskite semiconductors is attributed to the material's remarkably high defect tolerance and low exciton binding energy \cite{Yin2014feb,Walsh2015feb,Shi2015mara,Brandt2015jun,Emara2016jan}.
The theoretically predicted ionic defect formation energy is relatively low and consequently the equilibrium defect concentration should be quite high \cite{Kim2014apr,Walsh2015feb}.
For perovskite solar cells to reach their Shockley-Queisser limit, it is necessary to understand how these defects form and identify which ones contribute to non-radiative recombination, loss of photovoltage, and device hysteresis \cite{Stranks2017jul,Tress2018jan}.

At equilibrium, the concentration of a specific defect in a lead-halide perovskite crystal depends on the concentration (\latin{i.e.}, the chemical potential) of the relevant chemical species present in the solution or vapor from which the perovskite was precipitated \cite{Yin2014feb,Walsh2015feb,Shi2015mara,Emara2016jan,Senocrate2018aug}.
Nonequilibrium growth of the perovskite in the thin-film form \cite{Moore2015feb} should generate additional point- and grain-boundary defects.
The concentration of defects in the crystal also depends on the electron and hole chemical potential which --- if the perovskite's background carrier concentration is sufficiently low --- could be strongly affected by the substrate. 
Evidence that the substrate affects band alignment and induces p- or n-type conductivity can be seen in XPS \cite{Miller2014oct}, UPS \cite{Schulz2015may,Olthof2017jan}, and IPES \cite{Schulz2015may} measurements of lead-halide perovskite films, in one example in a film as thick as $\SI{400}{\nano\meter}$ \cite{Schulz2015may}.
How the substrate changes the near-surface and bulk conductivity of the perovskite is a topic of current research \cite{Olthof2017jan}; effects include the formation of an interface dipole, the creation of a chemically distinct passivation layer, and substrate-induced changes in perovskite film morphology.    

Defects in halide perovskites are challenging to study for a number of reasons.
Many of these defects are mobile under the application of electric field and/or light, with iodine species and vacancies considered to be most mobile species \cite{Ono2018mar,Yang2015jun,Delugas2016jul,Zhu2017aug,Li2016mar,Kim2018mar,Senocrate2017jun,Yuan2015aug,Futscher2018jan}.
Moreover, recent reports by Maier and coworkers show that the concentration of mobile iodine vacancies depends on illumination intensity \cite{Kim2018mar}.
This effect, which is expected from defect-energy calculations \cite{Yin2014feb,Shi2015mara,Senocrate2018aug}, needs to be considered in addition to the established effects of light on charge motion and polarization when trying to understand light-related hysteresis phenomena \cite{deQuilettes2016may,Zhao2015jul,Belisle2017jan,Yang2015jun,Senocrate2017jun,Wang2018feb,Pockett2015feb,Zarazua2016feb,Zarazua2016dec}.

Here we study a high-performing material with precursor solution stoichiometry \ce{(FA_{0.79}MA_{0.16}Cs_{0.05})_{0.97}Pb(I_{0.84}Br_{0.16})_{2.97}} (hereafter referred to as FAMACs) grown over four different substrates --- \ce{TiO2}, \ce{SnO2}, \ce{ITO}, and \ce{NiO}.\cite{Christians2018jan,Saliba2016jun,Draguta2018feb}
Christians and coworkers reported $1000$ hours \textit{operational} stability for FAMACs devices prepared with an \ce{SnO2} electron acceptor layer.\cite{Christians2018jan}
When compared to \ce{TiO2}-based devices, the \ce{SnO2} devices were much more stable.
While degradation of \ce{TiO2} devices has previously solely been attributed to ultraviolet light induced degradation, \cite{Domanski2017feb,Leijtens2013dec} they revealed, using ToF-SIMS measurements, different ionic distributions in \ce{TiO2}- and \ce{SnO2}-based devices after several hours of operation.
This observation demonstrates a clear difference in the light and/or electric field induced ion/vacancy motion in \ce{SnO2}- and \ce{TiO2}-based devices.

Motivated by these findings, here we measure the AC conductivity of the FAMACs films in the kHz to MHz regime and study this conductivity as a function of light intensity, time, and temperature.
We show that the light-on conductivity returns to its initial light-off value on two distinct timescales (sub $100$ ms and $10$'s of seconds) in the material grown on the electron accepting substrates \ce{TiO2} and \ce{SnO2}.
In contrast, material grown on the hole acceptor (NiO) and ITO substrates shows frequency-independent conductivity.
We tentatively assign these distinct behaviors to differences in the perovskites' background carrier type and concentration.
We find that the \ce{SnO2}-substrate films show higher dark conductivity and slower relaxation than the \ce{TiO2}-substrate films.
We show that at room temperature and above, the relaxation of the conductivity is activated over \ce{TiO2} (and possibly over \ce{SnO2} also).
Surprisingly, the relaxation of conductivity becomes considerably faster when the sample is cooled to a low temperature of \SI{233}{\K}.
The simplest explanation we can devise for these diverse observations is that the measured conductivity changes arise from light-dependent electronic fluctuations; at room temperature, these electronic fluctuations are coupled to slow, light-induced ionic/vacancy fluctuations that are frozen out at low temperature.
Our observation that the timescale of the conductivity recovery in the \ce{SnO2}-substrate sample is much slower than in the \ce{TiO2}-substrate sample supports the Christians \textit{et al.} hypothesis of slower ionic motion in the \ce{SnO2}-substrate sample compared to \ce{TiO2}-substrate sample \cite{Christians2018jan}.

These experiments were motivated by our previously reported scanning probe microscopy study of light- and time-dependent conductivity in a thin film of \ce{CsPbBr3} \cite{Tirmzi2017jan}.
We used sample-induced dissipation \cite{Denk1991oct,Stipe2001aug,Cockins2010may,Lekkala2012sep,Lekkala2013nov,Cox2016jun,Cox2013nov,Crider2006may,Stowe1999nov,Luria2012nov,Yazdanian2008jun,Hoepker2011oct,Kuehn2006apr,Kuehn2006aug} and broadband local dielectric spectroscopy (BLDS)\cite{Labardi2016may} to demonstrate for \ce{CsPbBr3} that conductivity shows a slow activated recovery when the light was switched off, with an activation energy and time-scale consistent with ion motion.
We concluded that the sample conductivity dynamics were controlled by the coupled motion of slow and fast charges.
While \ce{CsPbBr3} served as a sample robust to temperature- and light-induced degradation, it has a relatively high band gap and is thus poorly suited for use in high efficiency solar cells.
Many high efficiency devices reported to date rely on a mixed cation/anion perovskite absorber layer (such as FAMACs) to reach the desired bandgap and enhanced stability needed for photovoltaic applications.
The goal of the present study is to ascertain whether the conductivity dynamics observed for \ce{CsPbBr3} are evident in FAMACs films and to see whether they are substrate dependent.

As in Ref.~\citenum{Tirmzi2017jan}, here we follow conductivity dynamics using a charged microcantilever.
Microcantilevers are primarily used in scanning-probe microscope experiments to create images.
However, they have also proven useful in non-scanning experiments because of their tremendous sensitivity as force sensors.
Prior scanning probe microscopy (SPM) studies of lead-halide perovskite solar-cell materials have used Kelvin probe force microscopy to observe the dependence of the surface potential and surface photovoltage on time, electric field, and light intensity in order to draw conclusions about the spatial distribution of charges and ions \cite{Bergmann2014sep,Li2015dec,Yun2016jul,Yuan2017mar,Garrett2017mar,Collins2017aug,Salado2017sep,Xiao2017oct,Will2018jan,Birkhold2018may,Birkhold2018maya,Collins2018nov}.
In studies of organic solar cell materials, frequency-shift measurements have been used to follow the time evolution of photo-induced capacitance and correlate the photocapacitance risetime with device performance \cite{Coffey2006sep,Reid2010dec,Giridharagopal2012jan,Karatay2016may,Dwyer2017jun}.
Sample-induced dissipation has been used to monitor local dopant concentration in silicon \cite{Stowe1999nov} and GaAs \cite{Denk1991oct}; probing quantizied energy levels in quantum dots\cite{Cockins2010may};
examine photo-induced damage in organic solar cell materials \cite{Cox2013nov,Cox2016jun}; 
quantify local dielectric fluctuations in insulating polymers \cite{Kuehn2006apr,Kuehn2006aug,Crider2006may,Crider2007jul,Yazdanian2008jun,Hoepker2011oct};
and probe dielectric fluctuations and intra-carrier interactions in semiconducting small molecules \cite{Lekkala2012sep}.
Here we make use of the tremendous sensitivity of a charged microcantilever to passively observe the time evolution of a thin-film sample's conductivity \cite{Tirmzi2017jan} through changes in cantilever dissipation induced by conductivity-related electric-field fluctuations in the sample.\cite{Dwyer2018jul}


\section{Experimental Section}

\subsection{Materials}

Methylammonium bromide (\ce{CH3NH3Br}, \ce{MABr}), and formamidinium iodide (\ce{CH(NH2)2}, \ce{FAI}), were purchased from Dyesol and used as received. 
Lead (II) iodide ($99.9985\%$ metals basis) and the \ce{SnO2} colloid precursor (Tin(IV) oxide, $15\%$ in \ce{H2O} colloidal dispersion) were purchased from Alfa Aesar. 
All other chemicals were purchased from Sigma-Aldrich and used as received.

\subsection{Oxide Layer Deposition}
Indium tin oxide (ITO) glass was cleaned by sonication in acetone and isopropanol, followed by UV-ozone cleaning for $\SI{15}{\min}$.
Following cleaning, an additional thin oxide layer was deposited on the ITO glass (if necessary). 
\ce{TiO2} layers were deposited using a previously reported low temperature \ce{TiO2} process. 
Briefly, \ce{TiO2} nanoparticles were synthesized as reported previously by Wojciechowski \latin{et al.}\cite{Wojciechowski2014jan} and a $1.18$ wt. $\%$ ethanolic suspension, along with $\SI{20}{\mol}$ $\%$ titanium diisopropoxide bis(acetylacetonate), was spin-cast onto the ITO substrates with the following procedure: $700$ rpm, $\SI{10}{\sec}$; 1000 rpm, $\SI{10}{\sec}$; 2000 rpm, $\SI{30}{\sec}$. 
Tin oxide electron transport layers were deposited on cleaned ITO substrates.\cite{Jiang2016nov}
The aqueous \ce{SnO2} colloid solution, obtained from Alfa Aesar, was diluted in water with a ratio of $1:6.5$ and spin-cast at 3000 rpm for $\SI{30}{\sec}$. 
Both the \ce{TiO2} and \ce{SnO2} films were then dried at $\SI{150}{\celsius}$ for $30$ min and cleaned for $15$ min by UV-ozone immediately before use. 
NiO films were deposited from a solution of nixel nitrate hexahydrate and ethylenediamine in ethylene glycol following a previously reported procedure.\cite{Dunfield2018apr}

\subsection{FAMACs Perovskite Film Deposition}
Deposition of the FAMACs perovskite layers was carried out in a nitrogen glovebox following the method reported in Ref.~\citenum{Saliba2016jun}. 
The precursor solution was made by dissolving $\SI{172}{\milli\gram}$ FAI, $\SI{507}{\milli\gram}$ \ce{PbI2}, $\SI{22.4 }{\milli\gram}$ \ce{MABr}, and $\SI{73.4}{\milli\gram}$ \ce{PbBr2} ($1:1.1:0.2:0.2$ mole ratio) and $\SI{40}{\micro\liter}$ of \ce{CsI} stock solution ($\SI{1.5}{\molar}$ in DMSO) in $\SI{627}{\milli\gram}$ DMF and $\SI{183}{\milli\gram}$ DMSO ($4:1$ v/v).
The films were deposited by spin coating this precursor solution with the following procedure: $1000$ rpm for $\SI{10}{\sec}$, 6000 rpm for $\SI{20}{\sec}$. 
While the substrate was spinning, $\SI{0.1}{\milli\liter}$ of chlorobenzene was rapidly dripped onto the film with approximately $\SI{6}{\sec}$ remaining in the spin-coating procedure, forming a transparent orange film. The films were then annealed for $1$ hr at $\SI{100}{\celsius}$ to form highly specular FAMACs perovskite films.

\subsection{Scanning probe microscopy}

All experiments were performed under vacuum (\SI{5e-6}{\milli\bar}) in a custom-built scanning Kelvin probe microscope described in detail elsewhere \cite{Luria2011aug, Dwyer2017jun}.
The cantilever used was a MikroMasch HQ:NSC18/Pt conductive probe.
The resonance frequency and quality factor were obtained from ringdown measurements and found to be $\omega\st{c}/2\pi = f\st{c} = \SI{70.350}{\kilo\Hz}$ and $Q = \num{24000}$ respectively at room temperature.
The manufacturer's specified resonance frequency and spring constant were $f\st{c} = 60$ to $\SI{75}{\kilo\Hz}$ and $k = \SI{3.5}{\N\per\m}$.
Cantilever motion was detected using a fiber interferometer operating at $\SI{1490}{\nano\meter}$ (Corning model SMF-28 fiber).
More experimental details regarding the implementation of broadband local dielectric spectroscopy and other measurements can be found in the Supporting Information.


\section{Results}

\subsection{Theoretical background}
\label{Sec:theoretical-background}

\begin{figure*}[t]
    \includegraphics[width=6.50in]{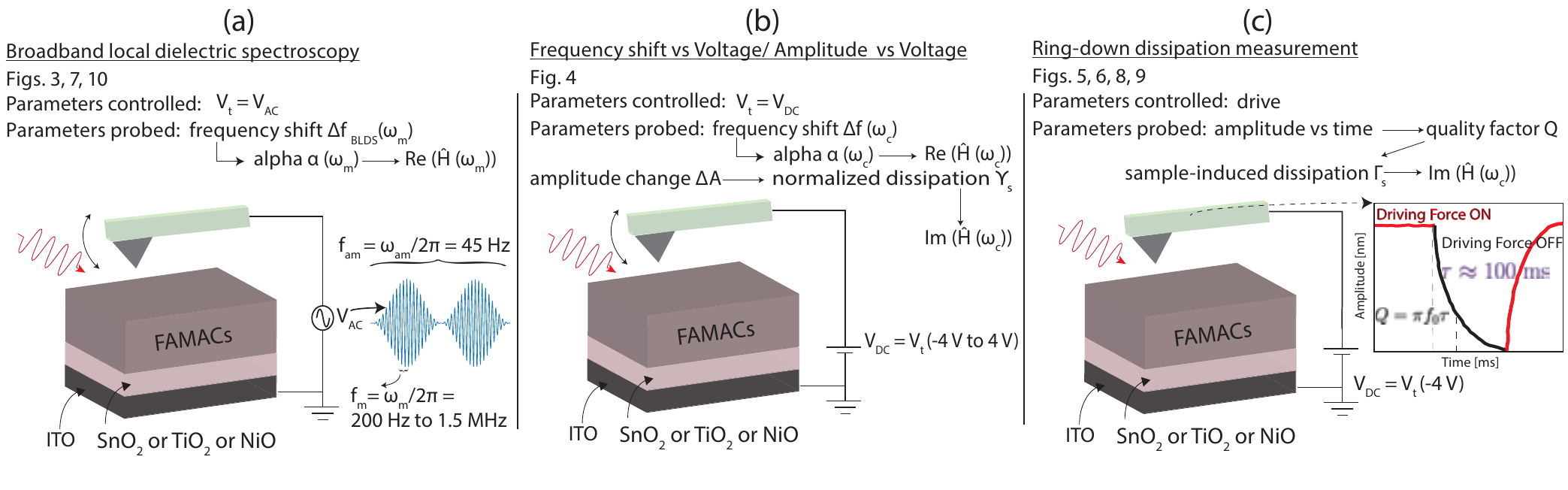}
    \caption{Schematic of the three scanning probe measurements used in this manuscript, highlighting the parameters that are controlled and probed. 
    	The measurements probe the complex frequency dependent tip-sample transfer function $\hat{H}$ (Eq.~\ref{eq:H} and Figure~\ref{fig:SI-equivalent-circuit-old-paper}).
		(a) In a broadband local dielectric spectroscopy measurement, the tip voltage $V\st{t} = V\st{AC}(\omega\st{m},\omega\st{am})$ is amplitude-modulated at a fixed frequency of $f\st{am} = \omega\st{am}/(2\pi) = \SI{45}{\hertz}$ and sinuodially modulated at a frequency $f\st{m} = \omega\st{m}/(2\pi)$ that is varied from $\SI{200}{\hertz}$ to $\SI{1.5}{M\hertz}$. 
		At each $f\st{m}$ the component of the cantilever frequency shift $\Delta f\st{BLDS} (\omega\st{m})$ at frequency $f\st{am}$ is measured using a frequency demodulator and lock-in amplifier and $\alpha(\omega\st{m})$ is calculated using Eq.~\ref{eq:alpha-kfc}. 
		The quantity $\alpha(\omega\st{m})$ primarily measures $\mathrm{Re}(\hat{H}(\omega\st{m}))$ (Eq.~\ref{eq:Deltaf-BLDS}).
		(b) In a frequency shift $\Delta f$ \latin{vs}.\ $V\st{t}$ measurement or an amplitude $A$ \latin{vs}.\ $V\st{t}$ measurement, the tip voltage $V\st{t} = V\st{DC}$ is slowly varied from $\SI{-4}{V}$ to $\SI{+4}{V}$ while the cantilever frequency shift $\Delta f$ and the cantilever amplitude $A$ are recorded.
		From $\Delta f$ a voltage-normalized frequency shift $\alpha(\omega\st{0})$ is calculated using Eq.~\ref{eq:fq-vs-voltage}; the quantity $\alpha(\omega\st{0})$ primarily measures $\mathrm{Re}(\hat{H}(\omega\st{c}))$. 
		From $A$ a voltage-normalized sample-induced dissipation $\gamma\st{s}$ is calculated using Eq.~\ref{eq:gamma-s-vs-voltage}; the quantity $\gamma\st{s}$ primarily measures $\mathrm{Im}(\hat{H}(\omega\st{c}))$ (Eqs.~\ref{eq:gamma-parob} and \ref{eq:Gamma-s}).
		(c) In a ring-down dissipation measurement, the cantilever drive is periodically switched off, the cantilever amplitude is measured as a function of time, and the cantilever's mechanical quality factor $Q$ is calculated from this ring-down transient.
		The observed change in $Q$ is converted to an equivalent change in sample-induced dissipation $\Gamma\st{s}$ using Eq.~\ref{eq:Gamma-t}.
		The parameter $\Gamma\st{s}$ primarily measures $\mathrm{Im}(\hat{H}(\omega\st{c}))$ (Eq.~\ref{eq:gamma-parob}).}
      \label{fig:Scheme-1}
\end{figure*}

Let us begin by summarizing the equations we will use to connect scanning-probe data to sample properties.  
Interested readers are directed to Ref.~\citenum{Tirmzi2017jan} and Ref.~\citenum{Dwyer2018jul} for a detailed derivation of the following equations.
In our measurements we modulate the charge on the cantilever tip and the sample and observe the resulting change in the cantilever frequency or amplitude.
This charge is modulated by physically oscillating the cantilever, by applying a time-dependent voltage to the cantilever tip, or by doing both simultaneously.
A summary of the distinct measurements carried out below is given in Figure~\ref{fig:Scheme-1}.

Changes in the cantilever frequency and amplitude may be expressed in terms of a transfer function $H$ which relates the voltage $V\st{ts}$ applied to the cantilever tip (the sample substrate is grounded) to the voltage $V\st{t}$ dropped between the cantilever tip and the sample surface. 
The cantilever is modeled electrically as a capacitor $C\st{tip}$ while the sample is modeled as resistor $R\st{s}$ operating in parallel with a capacitor $C\st{s}$ (Figure~\ref{fig:SI-equivalent-circuit-old-paper}).
The resulting transfer function is given in the frequency domain by 
\begin{equation}
\hat{H}(\omega) 
	= \frac{\hat{V}\st{t}(\omega)}{\hat{V}\st{ts}(\omega)} 
	= \frac{(j \omega C\st{tip})^{-1}}
	       {(R\st{s}^{-1} + j \omega C\st{s})^{-1} + (j \omega C\st{tip})^{-1}}
\label{eq:H-full}
\end{equation}
which simplifies to
\begin{equation}
\hat{H}(\omega)
    = \frac{1 + j \, g^{-1} \, \omega \tau\st{fast}}{1 + j \, \omega \tau\st{fast}}.
\label{eq:H}
\end{equation}
where $(R\st{s}^{-1} + j \omega C\st{s})^{-1}$ defines the sample impedance $Z$.
The transfer function $H$ can be viewed as a lag compensator whose time constant and gain parameter are given by, respectively,
\begin{equation}
\tau\st{fast} 
	= R\st{s} (C\st{s} + C\st{tip})
	= R\st{s} C\st{tot} 
	\hspace{0.5em} \text{and} \hspace{0.5em} 
g = C\st{tot} / C\st{s}.
\label{eq:tau-g}
\end{equation}
We give the time constant the subscript ``fast'' because of the time constant's similarity to ``$\tau\st{fast}$'' measured in impedance spectroscopy \cite{Tirmzi2017jan}.
We show experimentally below that $C\st{tip} \gg C\st{s}$; consequently, $\tau\st{fast} \approx R\st{s} C\st{tip}$.
This simplification allows us to associate photo-induced changes in cantilever frequency and amplitude to photo-induced changes in sample resistance or, equivalently, sample conductivity.

The complex-valued transfer function in Eq.~\ref{eq:H} has a real part which determines the in-phase forces and an imaginary part which determines the out-of-phase forces acting on the cantilever. 
We show the equivalent circuit and plot the shape of transfer function in Figure~\ref{fig:SI-equivalent-circuit-old-paper}.
The frequency shift measurements presented in Figure~\ref{fig:Dissipation-curvature}b probe the real part of the transfer function,
\begin{align}
\Delta f 
	& = -\frac{f\st{c}}{2k} \frac{\delta F^{\prime}}{A} \nonumber \\ 
	& = -\frac{f\st{c}}{4k}  \left(
		C_q^{\prime\prime} 
		+ \Delta C^{\prime\prime}
		\: \mathrm{Re} \left( \hat{H}(\omega\st{c}) \right)
	\right)
	\big (V - \phi \big)^2 
	\label{eq:f-parob}
\end{align}
where $f\st{c}= \omega\st{c}/2\pi$ is the resonance frequency, $k$ is the spring constant, and $A$ is the amplitude, respectively, of the cantilever; $F^{\prime}$ is the in-phase force; $C\st{t}$ is the cantilever capacitance computed at rest with the cantilever at its equilibrium position; $\Delta C^{\prime\prime} \equiv 2(C\st{t}^{\prime})^2/C\st{t}$, with primes indicating derivatives with respect the tip-sample distance; $\Delta C\st{q}^{\prime\prime} \equiv C\st{t}^{\prime\prime} - \Delta C^{\prime\prime}$; $V$ is the voltage applied to the cantilever tip; and $\phi$ is the surface potential.
The variable $\alpha\st{0}$ plotted in Figure~\ref{fig:Dissipation-curvature}a is a voltage-normalized frequency shift, the curvature of the $\Delta f$ \latin{vs}.\ $V$ data defined by the equation $\Delta f = \alpha_0 \, (V - \phi)^2$ and given by
\begin{equation}
\alpha\st{0} 
	= -\frac{f\st{c}}{4k}  \left(
		C_q^{\prime\prime} 
		+ \Delta C^{\prime\prime}
		\: \mathrm{Re} \left( \hat{H}(\omega\st{c}) \right)
	\right).
	\label{eq:alpha-parob}
\end{equation}
From Eq.~\ref{eq:alpha-parob} we can see that $\alpha\st{0}$ is sensitive to the real part of the transfer function at frequency $\omega\st{c}$, with additional contributions from in-phase forces present at low frequency ($\omega/2\pi < \SI{0.1}{\hertz})$.
The sample-induced dissipation plotted in Figures~\ref{fig:Dissipation-curvature}a, \ref{fig:dissipation-recovery-intensity}, \ref{fig:dissipation-recovery-temperature}, \ref{dissipation-tip-pulse}, and \ref{dissipation-recovery-cold-TiO-SnO} is sensitive to the out-of-phase part of the transfer function,
\begin{equation}
\Gamma\st{s} 
	= -\frac{1}{\omega\st{c}} \frac{\delta F^{\prime\prime}}{A} 
	= \frac{\Delta C^{\prime\prime}}{\omega\st{c}} 
	  \mathrm{Im} \left( \hat{H}(\omega\st{c}) \right) \big(V - \phi \big)^2,
\label{eq:gamma-parob}
\end{equation} 
where $F^{\prime\prime}$ is the out-of-phase force acting on the cantilever.
The voltage-normalized dissipation $\gamma\st{s} \propto \mathrm{Im}(\hat{H}(\omega\st{c}))$ plotted in Figure~\ref{fig:Dissipation-curvature}b is related to $\Gamma\st{s}$ through the equation $\Gamma\st{s} = \gamma\st{s}(V-\phi)^2$.
The BLDS measurements of Figures~\ref{fig:BLDS-roll-off}, \ref{fig:fixed-frequency-BLDS}, and \ref{BLDS-variable-temp} are frequency-shift measurements that probe the response of the sample to an oscillating applied voltage,
\begin{multline}
\Delta f\st{BLDS}(\omega\st{m}) =
-\frac{f\st{c} V\st{m}^2}{16 k} \Big[
C''_q + \Delta C'' \Re \Big( \hat{H}(\omega\st{m} + \omega\st{c}) \\
 + \hat{H}(\omega\st{m} - \omega\st{c}) \big) 
\Big]
\abs{\hat{H}(\omega\st{m})}^2
	\label{eq:Deltaf-BLDS}
\end{multline}
where $\omega\st{m}$ and $V\st{m}$ are the frequency and amplitude, respectively, of the oscillating applied voltage and we have assumed the amplitude modulation frequency $\omega\st{am}$ is much smaller than $\omega\st{m}$ (see Experimental Section in Supporting Information).
The imaginary part of the transfer function $\hat{H}$ is significant only at the frequency $1/\tau\st{fast}$ where the real part of the transfer function starts to roll-off.
The term in Eq.~\ref{eq:Deltaf-BLDS} containing the factors $\hat{H}(\omega\st{c} - \omega\st{m})$ and $\hat{H}(\omega\st{c} + \omega\st{m})$ is small as indicated by the BLDS spectra obtained at low light intensity over \ce{SnO2} and \ce{TiO2}, Figure~\ref{fig:BLDS-roll-off}, where the majority of the response rolls off at $\omega\st{m} \ll \omega\st{c}$.
We conclude that the BLDS measurement primarily measures the in-phase forces at the modulation frequency.
The voltage-normalized frequency shift $\alpha$ plotted in Figures~\ref{fig:BLDS-roll-off}, \ref{fig:fixed-frequency-BLDS}, \ref{BLDS-variable-temp} is related to $\Delta f\st{BLDS}$ by 
\begin{equation}
\alpha 
	= \frac{\Delta f\st{BLDS}(\omega\st{m})}{V\st{m}^{2}}.
	\label{eq:alpha-kfc}
\end{equation}
Figure~\ref{fig:Scheme-1} summarizes the experimental set-up and the measured quantity in each of the three different scanning probe measurements employed in this manuscript.


\subsection{Experimental findings}

\begin{figure}[t]
    \includegraphics[width=3.25in]{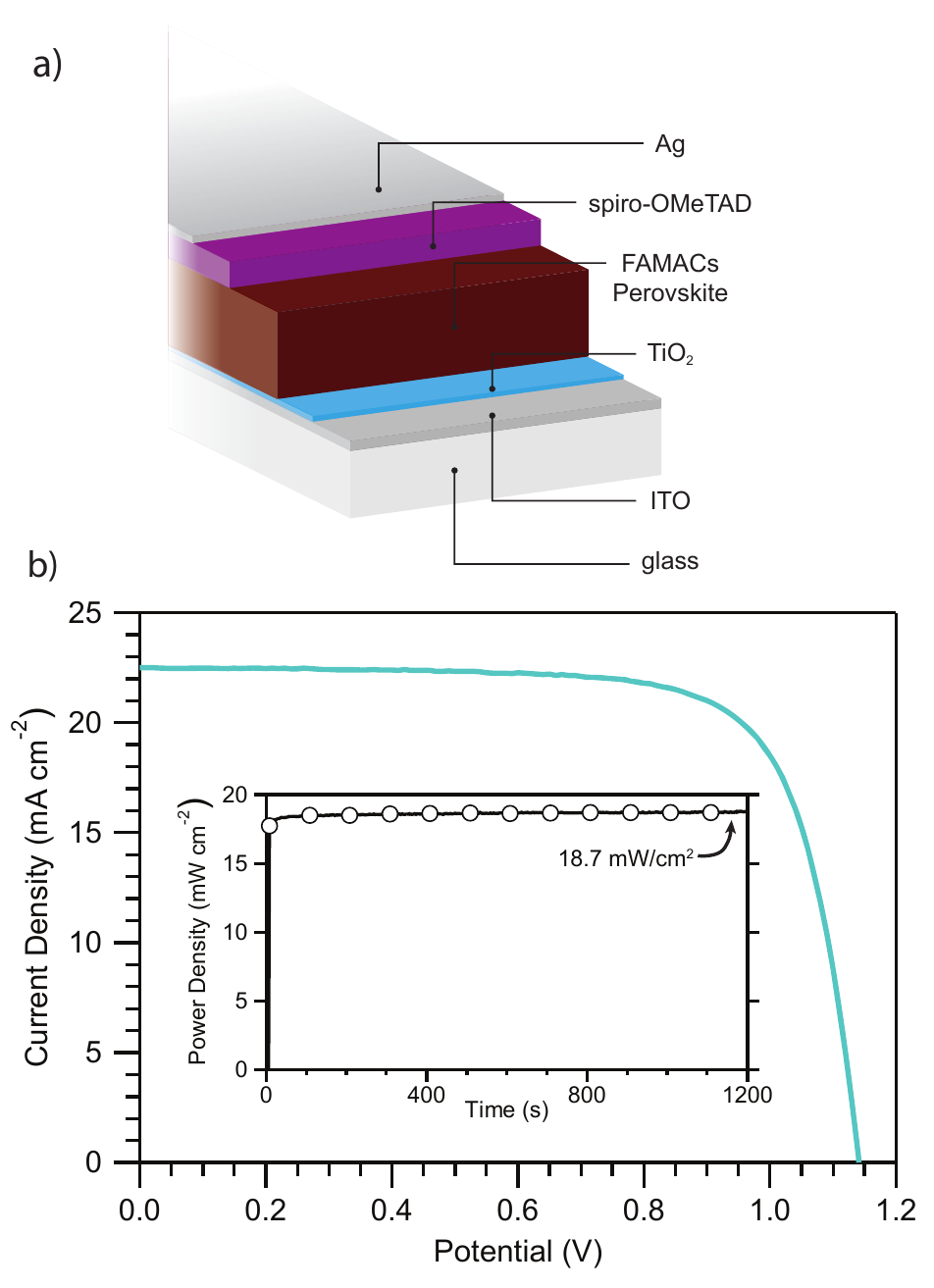}
    \caption{(a) Control perovskite solar cells were fabricated with the \ce{TiO2}/FAMACs/spiro-OMeTAD/Ag architecture depicted in the schematic. (b) A reverse scan (scan taken from open-circuit voltage, $V\st{OC}$, to short-circuit current, $J\st{SC}$) current density-voltage ($J-V$) curve taken for a device with the architecture depicted in (a). From the reverse $J-V$ scan, the $V\st{OC}$, $J\st{SC}$, fill factor, and power conversion efficiency were measured to be $\SI{1.141}{V}, \SI{22.49}{mA/cm^{2}}$, $0.744$, and $19.1\%$. The inset shows the power density of the device monitored at a constant bias of $\SI{0.94}{V}$ over the course of $\SI{1200}{s}$ which was found to stabilize at $18.7\%$ efficiency.}
    \label{fig:Device}
\end{figure}

We now present data acquired on the FAMACs samples prepared on a range of substrates.
All of the substrates (\ce{TiO2}, \ce{SnO2}, \ce{NiO}, \ce{ITO}) are planar structures.
The FAMACs thickness was ${\sim}\SI{700}{\nano\meter}$; the thickness of ETL/HTL layer was ${\sim}\SI{40}{\nano\meter}$ with average roughness of ${\sim}\SI{10}{\nano\meter}$, and the thickness of the ITO was ${\sim}\SI{100}{\nano\meter}$ with an average roughness of ${\sim}\SI{2}{\nano\meter}$.
The samples were illuminated from the top.
The high absorption coefficient of the perovskite film means that electron and hole generation was confined to the top ${\sim}\SI{200}{\nano\meter}$ of the sample, a distance significantly smaller than the $\SI{700}{\nano\meter}$ thickness of the FAMACs layer.
Figure~\ref{fig:Device} shows device-performance data for a representative FAMACs film prepared on a \ce{TiO2} substrate; this data demonstrates the high quality of the films used in this study.
\begin{figure*}[t]
    \includegraphics[width=5.75in]{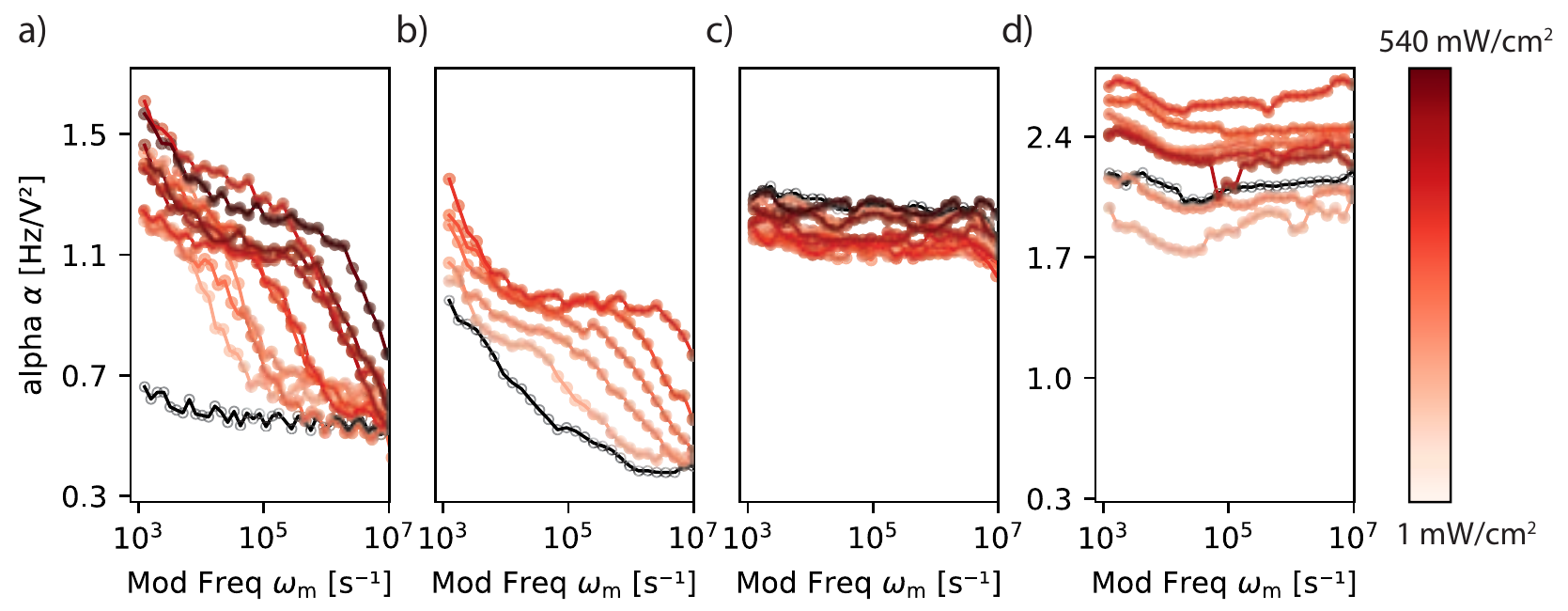}
    \caption{Distinct Broadband Local Dielectric Spectroscopy (BLDS) curves seen in FAMACs thin films grown on 
    (a) \ce{TiO2},
    (b) \ce{SnO2},
    (c) \ce{ITO}, and
    (d) \ce{NiO}.
    Curves are colored according to the applied light intensity (See right hand legend)and vertically opffset by 0.05 for clarity. For reference, BLDS curve in the dark is shown in open black circles. Experimental parameters: modulation voltage $V\st{m} = \SI{6}{\V}$, $\lambda =$ $\SI{639}{\nano\meter}$, tip-sample separation $h = \SI{200}{n\m}$ except for the \ce{TiO2}-substrate sample where $h = \SI{150}{n\m}$.}
    \label{fig:BLDS-roll-off}
\end{figure*}

Figure~\ref{fig:BLDS-roll-off} shows Broadband Local Dielectric Spectroscopy data (BLDS \cite{Labardi2016may}, Fig.~\ref{fig:Scheme-1}a) acquired of films prepared on \ce{TiO2}, \ce{SnO2}, \ce{ITO}, and \ce{NiO} substrates (see the Experimental Section in Supporting Information).
In Figure~\ref{fig:BLDS-roll-off}, a decrease in $\alpha$ at large voltage-modulation frequency $\omega\st{m}$ indicates qualitatively that not all of the sample charge is able to follow the modulated tip charge.
In our impedance model of the tip-sample interaction, Figure~\ref{fig:SI-equivalent-circuit-old-paper}, this decrease is attributed to the $RC$ roll-off of the tip-sample circuit.
A light-dependent change in the roll-off frequency is consistent with sample conductivity increasing with increasing light intensity or, in other words, a decrease in the time constant $\tau\st{fast}$ with light.
In Figure~\ref{fig:BLDS-roll-off} we clearly see a roll-off of the $\alpha$ \latin{vs}.\ $\omega\st{m}$ curves that depends on the light intensity in the case of electron-acceptor substrates (\ce{TiO2} and \ce{SnO2}), whereas in the case of the \ce{NiO}-(hole acceptor) and \ce{ITO}-substrate samples, $\alpha$ is independent of both $\omega\st{m}$ and light intensity.

For the rest of this section of the manuscript, we compare the light and frequency dependence of the conductivity in the \ce{TiO2} and \ce{SnO2}-substrate samples.
Both samples show a light-dependent roll-off of the dielectric response.
However some significant differences can also be seen:
\begin{enumerate}
\item In the dark, the \ce{SnO2}-substrate sample is more conductive than the \ce{TiO2}-substrate sample as seen by their dark BLDS response curves.
\item The conductivity of the \ce{SnO2}-substrate sample is more strongly affected by light than that of the \ce{TiO2}-substrate sample; the roll-off moves to higher frequencies for the same light intensity for the \ce{SnO2}-substrate sample compared to the \ce{TiO2}-substrate sample.
\end{enumerate}

\begin{figure}[t]
  \includegraphics[width=2.50in]{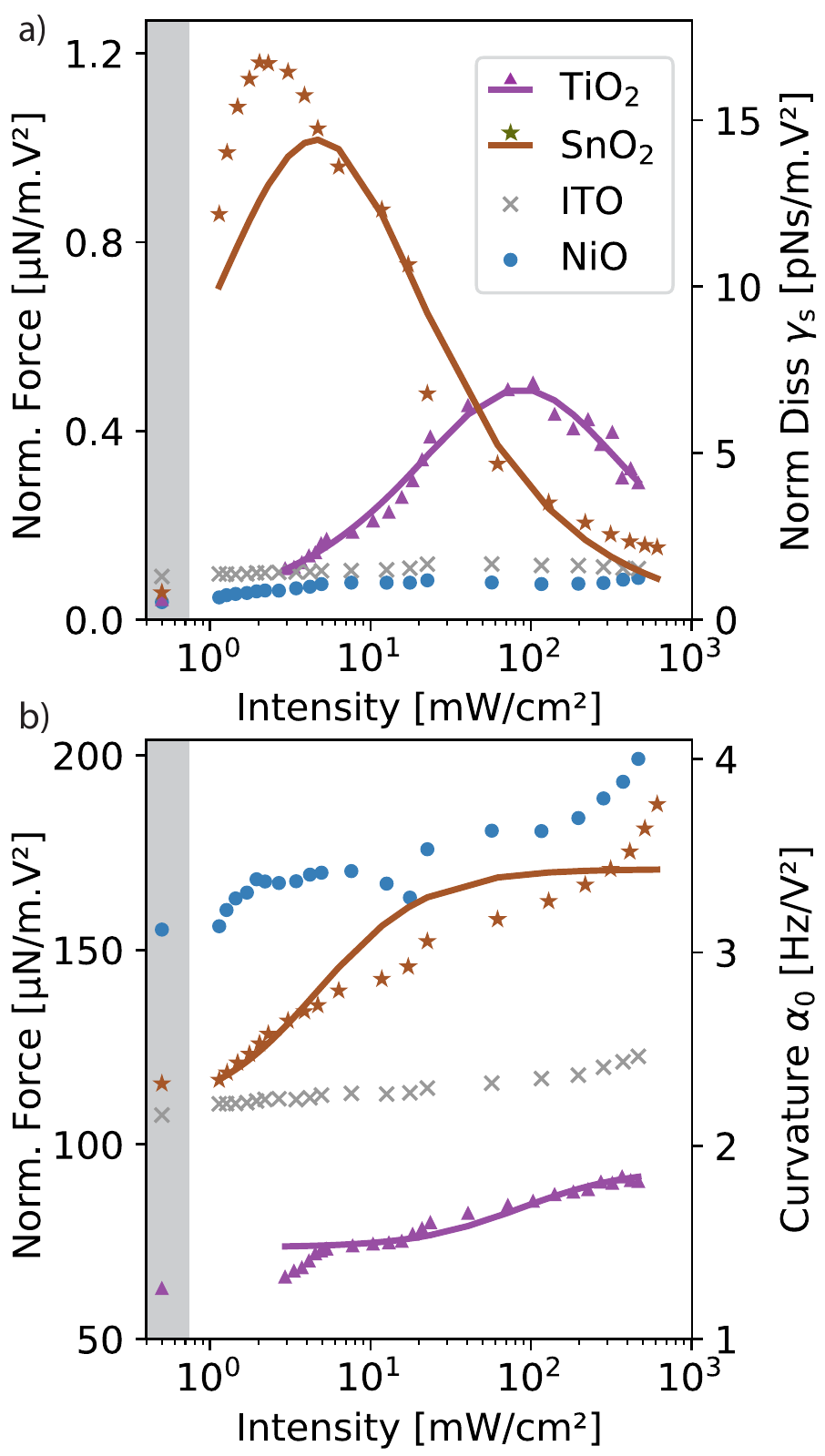}
  \caption{Normalized 
    (a) out-of-phase (dissipation) 
    (b) in-phase (curvature) force for FAMACs films on different substrates.
    Shaded points indicate values in the dark.
    Solid lines show a fit to the impedance model described in Ref~\citenum{Tirmzi2017jan}.
    Experimental parameters: $\lambda = \SI{639}{\nano\meter}$, 
    $h = \SI{200}{\nano\meter}$, 
    $A\st{0} = \SI{45}{\nano\meter}$, 
    $V\st{ts} = \SI{-4}{\V}$ to $\SI{4}{\V}$ (bipolar sweeps).}
  \label{fig:Dissipation-curvature}
\end{figure}

These light-dependent conductivity effects can be confirmed through quasi-steady-state measurements of the cantilever frequency shift ($\Delta f$) \latin{vs}.\ applied tip voltage ($V\st{ts}$) and cantilever amplitude ($A$) \latin{vs}.\ applied tip voltage (Figure~\ref{fig:Scheme-1}b).
In these measurements the cantilever is driven using constant-amplitude resonant excitation and the cantilever amplitude and frequency shift are recorded at each applied $V\st{ts}$.
By fitting the measured frequency shift and amplitude data to Eq.~\ref{eq:fq-vs-voltage} and Eq.~\ref{eq:gamma-s-vs-voltage} respectively --- see Figure~\ref{fig:SI-Amplitude-Freq-example} for representative curves and Sec.~\ref{sec:curvature-dissipation-calculation} for calculation details --- we can calculate the curvature ($\alpha\st{0}$) change and a voltage-normalized sample-induced dissipation constant ($\gamma\st{s}$).
These values are not affected by the tip voltage sweep width; the large wait time ($\SI{500}{\milli\second}$) employed at each applied tip voltage ensure that the measured response is a steady-state response.

Figure~\ref{fig:Dissipation-curvature}a shows the sample-induced dissipation measured through the amplitude-voltage method. 
In this measurement we are sensitive to the out-of-phase response of the sample at the cantilever frequency. 
Figure~\ref{fig:cartoon-old-paper} illustrates the predicted dependence of the curvature and sample-induced dissipation on light intensity.
A non-linear increase in dissipation which reaches a maximum and then decreases with light intensity can be explained by the existence of a time constant $\tau\st{fast}$ that increased monotonically with light intensity.
At high light intensities, the sample reaches its high-conductivity state.
When $\tau\st{fast}$ is less than $\omega\st{c}^{-1} = \SI{2}{\micro\second}$, most of the sample charge responds instantaneously to changes in the tip position, leading to a decrease in the out-of-phase force acting on the cantilever and a reduction in sample-induced dissipation.
We see for both the \ce{TiO2}-substrate sample and the \ce{SnO2}-substrate sample that dissipation reached a maximum before decreasing when the light intensity was increased monotonically.
Figure~\ref{fig:Dissipation-curvature}b shows that, concomitant with a dissipation peak, there is a non-linear change in the in-phase response at the cantilever frequency, observed as a changes in the curvature of the frequency shift \latin{vs}.\ applied tip voltage parabola ($\alpha\st{0}$). 

The data of Figure~\ref{fig:Dissipation-curvature}, which primarily measures sample response at a single frequency ($\omega\st{c}$), corroborates the data of Figure~\ref{fig:BLDS-roll-off} which shows the sample response at multiple frequencies.
The solid lines in Figure~\ref{fig:Dissipation-curvature} are a fit to a one-time-constant impedance model described in Ref.~\citenum{Tirmzi2017jan} and summarized in Sec.~\ref{Sec:theoretical-background}.
The model \emph{qualitatively} explains the seemingly-anomalous peak in sample-induced dissipation \latin{vs}.\ light intensity data over both \ce{TiO2} and \ce{SnO2}.
The one-time-constant model only qualitatively describes the charge dynamics in the \ce{SnO2}-substrate sample; adding further electrical components to the sample-impedance model, justified by the double roll-off seen in the Figure~\ref{fig:BLDS-roll-off}b data, would improve the \ce{SnO2}-substrate fits in Figure~\ref{fig:Dissipation-curvature}.
\begin{figure}[t]
\includegraphics[width=3.25in]{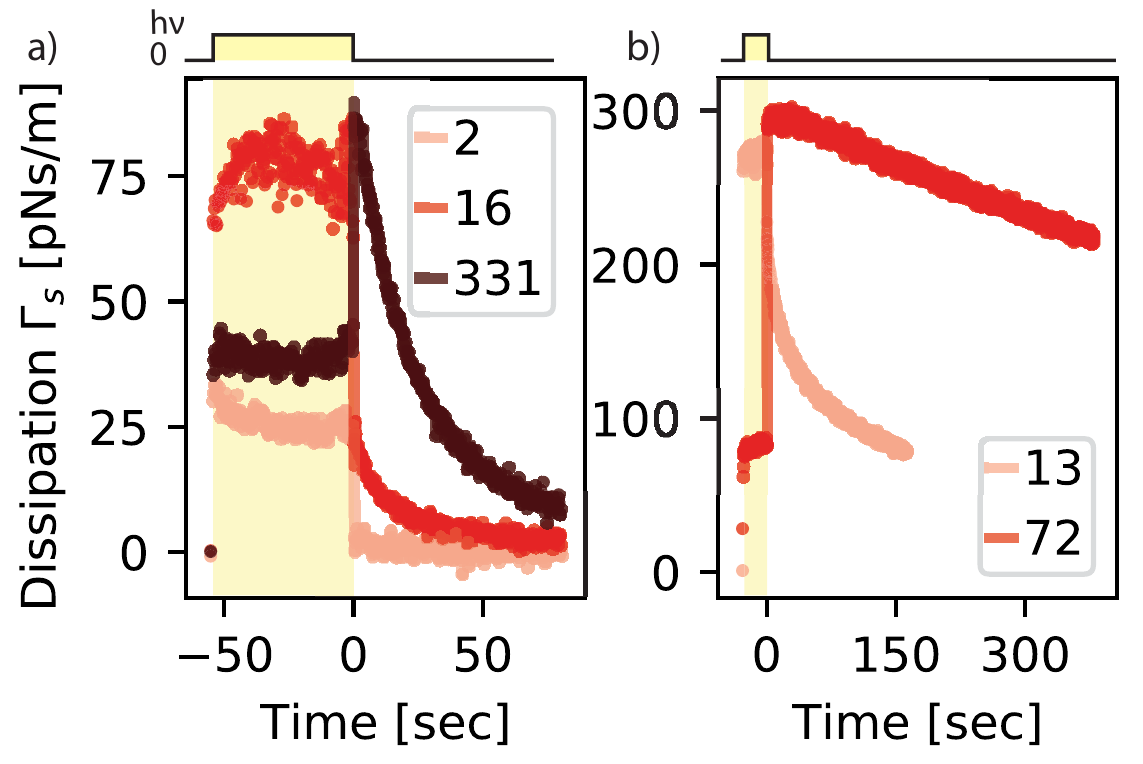}
\caption{Dissipation recovery in the dark is substrate dependent.
Dissipation \latin{vs}.\ time for FAMACs film on (a) \ce{TiO2} (b) \ce{SnO2}-substrate. 
The indicated illumination intensity was turned off at $t = \SI{0}{sec}$. Experimental parameters: $V\st{ts} = \SI{-4}{\V}$, $T = \SI{292}{\K}$, (a) $\lambda =$ $\SI{535}{\nano\meter}$, $h = \SI{175}{\nano\meter}$, (b) $\lambda =$ $\SI{639}{\nano\meter}$,  $h = \SI{200}{\nano\meter}$. }  
         \label{fig:dissipation-recovery-intensity}
\end{figure}
 
We observed the dynamics of $\tau\st{fast}$ in real time through two different methods.
In Figure~\ref{fig:dissipation-recovery-intensity}a, we show how the dissipation changes for the \ce{TiO2}-substrate sample for different light intensities. 
Here we inferred sample-induced dissipation by measuring changes in the quality factor of the cantilever through a ring-down measurement (Figure~\ref{fig:Scheme-1}c and Sec.~\ref*{sec:Q-recovery}).
The recovery of dissipation clearly had two distinct timescales --- a fast component and a slow component. 
In Figures~\ref{fig:dissipation-recovery-intensity}a and b, when the light was switched on, there was a large and prompt ($\leq$ $\SI{100}{\milli\second}$) increase in dissipation followed by a small and much slower increase that lasted for $\SI{10}{\second}$ or longer.
The presence of the slow component was especially clear when the light intensity was greater than the light intensity giving the maximum dissipation.
Whether the dissipation increased or decreased when the light was switched off depended on the value that $\tau\st{fast}$ (\latin{i.e.}\ sample conductivity) reached during the light-on period.
At low light intensities ($\SI{2.06}{\milli\W\per\cm\squared}$ and $\SI{15.6}{\milli\W\per\cm\squared}$ for the \ce{TiO2}-substrate sample,  Figure~\ref{fig:dissipation-recovery-intensity}a, and $\SI{13}{\milli\W\per\cm\squared}$ for the \ce{SnO2}-substrate sample, Figure~\ref{fig:dissipation-recovery-intensity}b), the dissipation $\Gamma\st{s}$ decreased when the light was switched off, indicating that $\tau\st{fast}$ was $\geq \SI{2}{\micro\second}$.
On the other hand, the initial rise in $\Gamma\st{s}$ when the light was switched off for the $\SI{331}{\milli\W\per\cm\squared}$ dataset in Figure~\ref{fig:dissipation-recovery-intensity}a and the $\SI{72}{\milli\W\per\cm\squared}$ dataset in Figure~\ref{fig:dissipation-recovery-intensity}b is consistent with a light-on $\tau\st{fast}$ being $\leq \SI{2}{\micro\second}$.
In such a case, when the light was switched off, the $\Gamma\st{s}$ promptly increased in $\leq \SI{200}{\milli\second}$ as $\tau\st{fast}$ approached the value of $\SI{2}{\micro\second}$. 
Subsequently, $\Gamma\st{s}$ gradually decreased over $10$'s of seconds as $\tau\st{fast}$ became $\geq \SI{2}{\micro\second}$.

In Figure~\ref{fig:dissipation-recovery-intensity}b, we show a time-resolved light-induced dissipation measurement for the \ce{SnO2}-substrate sample. 
The slow part of the recovery of dissipation was extremely slow ($>$ $\SI{500}{\sec}$) at room temperature.
This slow recovery indicates that the \ce{SnO2}-substrate sample retained its conductive state for a much a longer time than did the \ce{TiO2}-substrate sample. 
Interestingly, the slow time constant for dissipation recovery showed a dependence on the pre-soak intensity.
As the light intensity increased, the dissipation recovery \textit{time constant} became slower.
The underlying process responsible for the dissipation recovery is thus light-intensity dependent.
The prompt recovery in conductivity was too fast to resolve, limited by the $\SI{100}{\ms}$ time resolution of the ringdown measurement.
\begin{figure}[t]
\includegraphics[width=3.25in]{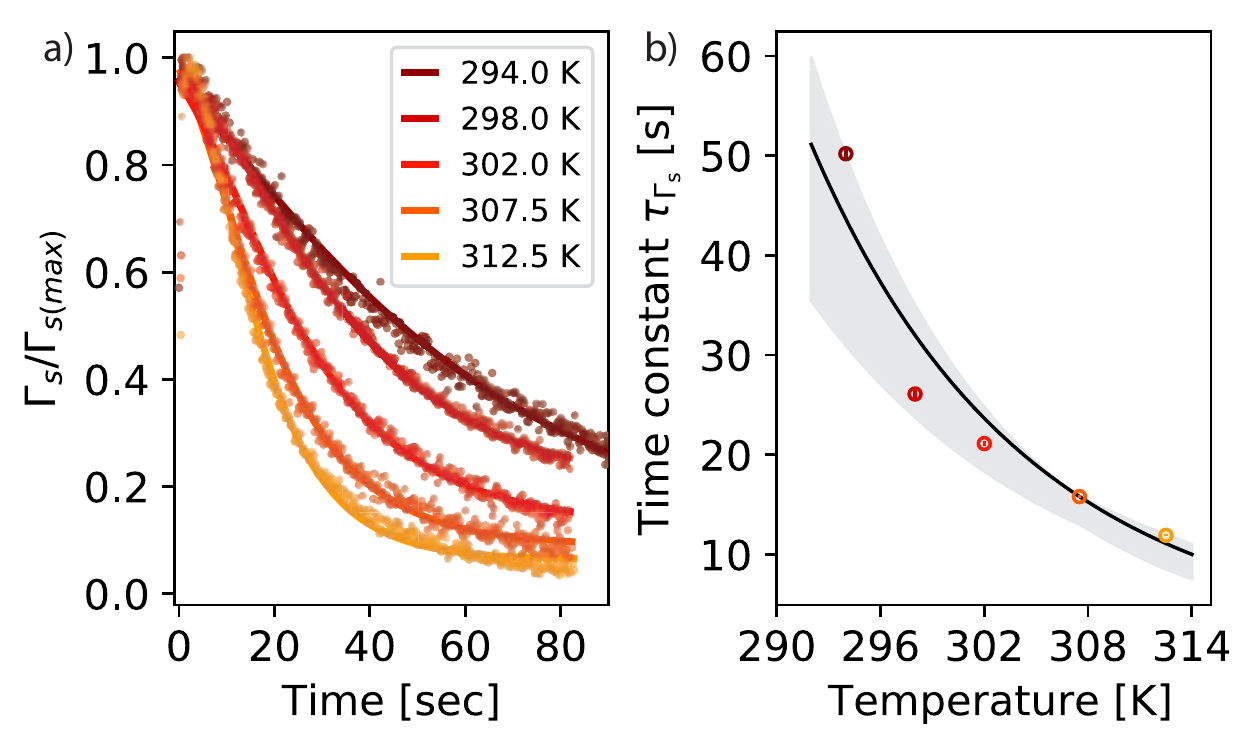}
         \caption{Dissipation recovery over the \ce{TiO2}-substrate sample is temperature dependent. 
        (a) Dissipation recovery transients in the dark after a period of continuous illumination.
        (b) Time constant for dissipation recovery $\Gamma_{\st{s}}$ recovery \latin{vs}.\ temperature with error bars          calculated from the fits in (a). The shaded region represents $2$ standard deviations for a weighted least squares fit to an exponential model $\tau\st{\Gamma}(T) = A^{-1} \exp (E\st{a} / k\st{B} T)$.
    The best fit parameters with two standard deviation error bars were $E\st{a} = \SI{0.58}{} \pm \SI{0.07}{\eV}$ and $A = \SI{4.9}{} \pm \SI{1.0e2}{\per\s}$.
    Experimental parameters: $V\st{ts} = \SI{-4}{\V}$, $I\st{h\nu} = \SI{292}{\milli\watt/\cm\squared}$, $t\st{soak} = \SI{27}{\second}$, $h = \SI{150}{\nano\meter}$, $\lambda =$ $\SI{639}{\nano\meter}$.}
         \label{fig:dissipation-recovery-temperature}     
\end{figure}
\begin{figure*}[t]
\includegraphics[width=4.50in]{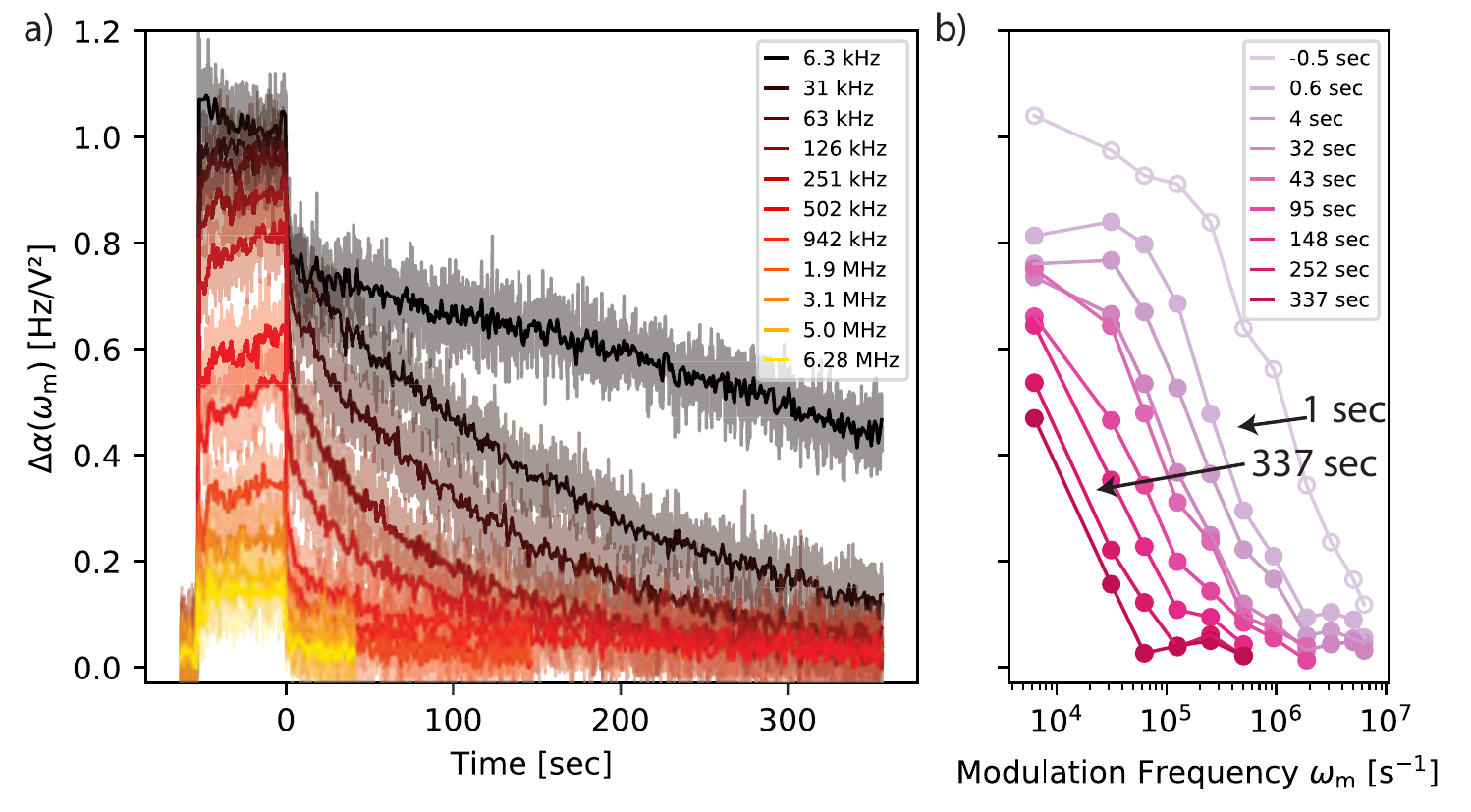}
\caption{Dielectric response for the \ce{TiO2}-substrate sample recovers on multiple timescales.
See legend for modulation frequency.
(a) Dielectric response \latin{vs}.\ time measured at the give, fixed modulation frequency $\omega_{\st{m}}$. Large bandwidth (shaded) and $\SI{1}{\sec}$ running average (solid line).
(b) Deduced dielectric response curves at various time offsets from the data in (a). 
Experimental parameters: $I\st{h\nu} =\SI{292}{\milli\W/\cm\squared}$ , $h = \SI{150}{\nano\meter}$, $V\st{m} = \SI{4}{\V}$, $\lambda =$ $\SI{639}{\nano\meter}$.
}
\label{fig:fixed-frequency-BLDS}
\end{figure*}

In Figure~\ref{fig:dissipation-recovery-temperature}a we show the temperature dependence of the slow part of the dissipation recovery. 
In Figure~\ref{fig:dissipation-recovery-temperature}b we plot the calculated dissipation-recovery time constant ($\tau\st{\Gamma}$) for the data in (a) versus temperature.
The slow part of the dissipation recovery is activated with an activation energy $E\st{a} = 0.58 \pm \SI{0.07}{\eV}$ for the \ce{TiO2}-substrate sample.
For the \ce{SnO2}-substrate sample, the slow part of the dissipation recovery did  not show an appreciable change in the accessible $\SI{300}{\K}$ to $\SI{315}{\K}$ temperature window (Figure~\ref{SnO-high-temp}).
The dissipation recovery of the \ce{SnO2}-substrate sample was much slower than \ce{TiO2}-substrate sample at room temperature, implying $E\st{a} \geq$ $\SI{0.58}{\eV}$.

To further show the presence of two distinct recovery timescales, we examined the fixed-frequency dielectric response for the \ce{TiO2}-substrate sample (Figure~\ref{fig:fixed-frequency-BLDS}). 
Here we illuminate the sample and measure the time-resolved dielectric response at a fixed modulation frequency ($\omega\st{m}$).
The response at each $\omega\st{m}$ corresponds to the in-phase force at that modulation frequency. 
By doing the measurement at different modulation frequencies, we can visualize the time evolution of the full dielectric response curve in the dark after the light was turned off.
Comparing the reconstructed dielectric response curves for time just before switching off the light (open circles) and $\SI{1.1}{\sec}$ after switching off the light (closed circles) shows a fast ($<$ $\SI{1}{\sec}$) decrease in the roll-off frequency (\latin{i.e.} a decrease in sample conductivity). 
This fast decrease was followed by a slow decrease lasting $100$'s of seconds before the dark state is reached.
Thus sample conductivity was thus decreasing on multiple distinct timescales in the dark.
While the dielectric response curve measurement produces a more comprehensive picture of the conductivity recovery compared to the single-shot ring-down measurements, it is an inherently slower measurement than the ring-down measurement and is potentially affected by hysteresis since it requires a long resting time ($\SI{600}{\second}$ between each measurement).

We attribute changes in dissipation and the BLDS response to changes in sample resistance $R\st{s}$ (or conductivity) rather than sample capacitance $C\st{s}$.
In Figures~\ref{fig:BLDS-roll-off}a,b and Figure~\ref{fig:fixed-frequency-BLDS}, the high frequency response, determined by the ratio $C\st{t}/(C\st{s}+C\st{t})$, is independent of the light intensity.
The similar high frequency response implies that $C\st{s}$ does not depend strongly on the light intensity.
Therefore we can make the approximation that changes in the BLDS response primarily reflect changes in the sample resistance, or equivalently, sample conductivity.
In a mixed ionic-electronic conductor, one might expect $R\st{s}$ to report on changes in the ambipolar resistance (or conductivity) and the measured dynamics of the resistance changes would be determined by the slowest diffusing species \cite{Kerner2017dec}.
Using a more accurate transmission-line model of sample impedance, we show below in Sec.~\ref{Sec:ambipolar-conductivity} that our electric force microscope measurements are probing the \emph{total} sample conductivity (Eqs.~\ref{Eq:H-approx}).

\begin{figure}[t]
\includegraphics[width=3.25in]{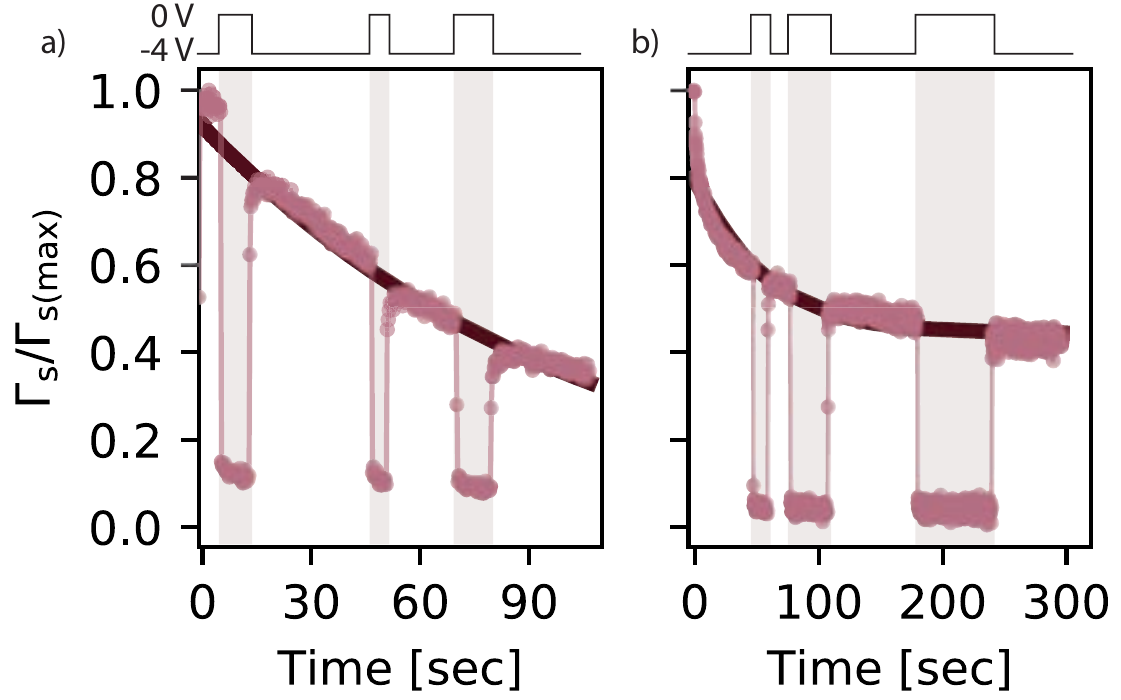}
\caption[Tip electric field effect on dissipation transient]{Recovery of the dissipation in the dark is unaffected by tip voltage $V\st{ts}$ for FAMACs films on (a) \ce{TiO2} and  (b) \ce{SnO2}.
Tip voltage is turned to $V\st{ts} = \SI{0}{\V}$ for different duration of time during dissipation recovery. Solid lines are fit to a simple exponential model for dissipation recovery.
        Experimental parameters: (a) pre-soak $I_{h\nu} = \SI{292}{\milli\W\per\cm\squared}$, $V\st{ts} = \SI{-4}{\V}$, $h = \SI{150}{\nano\meter}$) (b) pre-soak $I_{h\nu} = \SI{5}{\milli\W\per\cm\squared}$, $\lambda =$ $\SI{639}{\nano\meter}$, $V\st{ts} = \SI{-4}{\V}$, $h = \SI{200}{\nano\meter}$), typical $\Gamma\st{max} = \SI{100}{pNs/m}$. }
\label{dissipation-tip-pulse}
\end{figure}

To verify that the tip electric field is not the cause of the slow dissipation recovery, we switched off the tip voltage during acquisition of the dissipation recovery transients for different durations of time (Figure~\ref{dissipation-tip-pulse}). 
We find negligible differences in the dissipation recovery transient when the tip voltage is switched off.
We conclude that we are passively observing fluctuations in the sample whose dynamics are unaffected by the tip charge,  \latin{i.e.} we are operating in the linear-response regime of fluctuation-dissipation theorem.
This finding indicates that $\tau\st{fast}$ represents sample conductivity that continues to relax irrespective of the tip electric field at the surface.
This experimental result also rules out that changes in the conductivity are due to tip-induced charging and discharging of the interfacial redistribution of electronic and ionic charges at the perovskite-substrate interface.
\begin{figure}[t]
\includegraphics[width=3.25in]{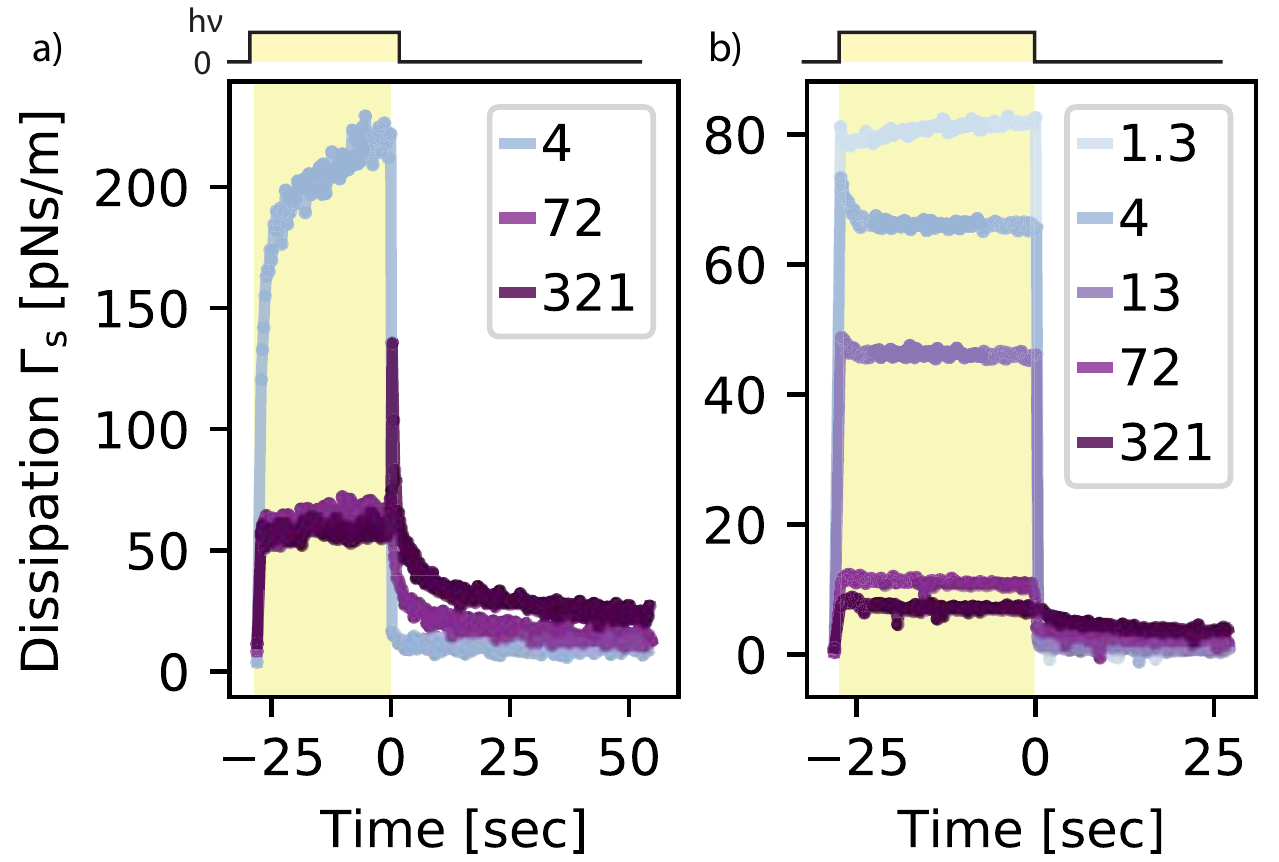}
\caption{Slow dynamics of dissipation recovery are suppressed at 233 K.
Low temperature dissipation recovery  at selected light intensities ($\lambda =$ $\SI{639}{\nano\meter}$) for FAMACs film on (a) \ce{TiO2} and (b) \ce{SnO2}.}
\label{dissipation-recovery-cold-TiO-SnO}
\end{figure}

Lead halide perovskites are worse thermal conductors compared to many organic semiconductors and at normal solar cell operating conditions, thermal-gradient-induced ion migration away from the light source due to the Soret effect is a possibility.\cite{Walsh2018jul}. 
However, we can rule out temperature variations induced due to light as the main cause of the slow dissipation recovery.
The slow recovery is evident at even very modest light intensity of $\SI{13}{{\milli\W\per\cm\squared}}$ in Figure~\ref{fig:dissipation-recovery-intensity}a and $\SI{15}{{\milli\W\per\cm\squared}}$ in Figure~\ref{fig:dissipation-recovery-intensity}b.
Following the analysis of photo thermal effects presented in Ref.~\citenum{Dwyer2017jun}, even $\SI{300}{{\milli\W\per\cm\squared}}$ would cause $< \SI{1}{K}$ change in temperature.
Additional analysis of the data presented in Figure~\ref{fig:BLDS-roll-off} is provided in Figure~\ref{TiO-time-constant} and Figure~\ref{SnO-time-constant} and shows that $\tau\st{fast}$ essentially decreases logarithmically with light intensity $I$ ($\tau\st{fast} \propto (\log{I})^{-1}$).
Photothermal effects would be inconsistent with this experimental result. 
\begin{figure}[t]
\includegraphics[width=2.50in]{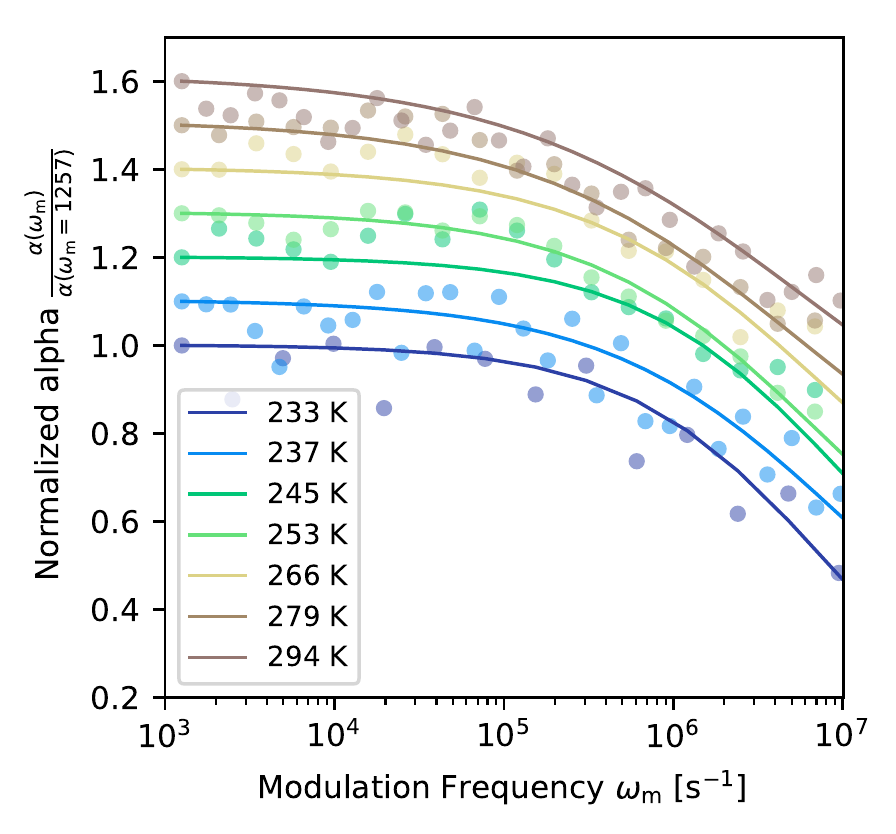}
\caption{Qualitatively similar roll-off frequency for the BLDS curves shows minimal effect of temperature on the light dependent conductivity. 
Normalized BLDS spectra for \ce{SnO2}-substrate sample taken at various temperatures at the same light intensity ($I\st{h\nu} =\SI{13}{\milli\W/\cm\squared}$, $\lambda =$ $\SI{639}{\nano\meter}$) during gradual heating from $T = \SI{233}{\K}$ to $T = \SI{294}{\K}$ offset by $0.1$. Solid lines are a fit to a one time constant low pass filter and are presented as a guide only.}
\label{BLDS-variable-temp}
\end{figure}

We next measure the effect of significantly reduced ion motion on the dynamics of sample conductivity.
Several reports suggest that, in a similar temperature range (\SI{233}{\K}), the effect of ion motion on measurements can be significantly reduced or eliminated by cooling the sample.
This reduction manifests itself in device measurements as a reduced hysterisis in $JV$ curves.\cite{Ginting2016sep,Zou2016apr}
This motion-reduction hypothesis was investigated here by measuring dissipation recovery dynamics at low temperature (\SI{233}{\K}). 
In Figure~\ref{dissipation-recovery-cold-TiO-SnO} we see that the dissipation \latin{vs}.\ light intensity showed a similar behavior to the room temperature measurements of Figure~\ref{fig:dissipation-recovery-intensity} for the duration of illumination.
This finding is further corroborated by BLDS measurements at a fixed light intensity taken at different temperatures for the \ce{SnO2}-substrate sample.
This data shows that the value of $\tau\st{fast}$ in \ce{SnO2} is unaffected by temperature (Figure~\ref{BLDS-variable-temp}) and $\tau\st{fast}$ is decreased with increasing light intensity even as the temperature is lowered.
Interestingly, there is essentially no slow recovery of dissipation when compared with room temperature (Figure~\ref{dissipation-recovery-cold-TiO-SnO}).
The absence of slow recovery dynamics is consistent with the hypothesis that the slow recovery (Figure~\ref{dissipation-recovery-cold-TiO-SnO}) is determined by ion motion, which is substantially arrested at \SI{233}{\K}.
The total conductivity of the sample under illumination is dominated by the electronic carriers.\cite{Kim2018mar}
The interaction of these electronic carriers with the slow moving ions determines the dynamics of the light-induced conductivity decrease when the light is switched off.


\section{Discussion}

\subsection{Summary of 
  experimental conductivity findings}

We have observed photo-induced changes in conductivity perturbing the electrostatic forces oscillating both in-phase and out-of-phase with the cantilever motion. 
Light-induced changes in the in-phase force leads to the frequency-shift effects seen in Figures~\ref{fig:BLDS-roll-off}, \ref{fig:Dissipation-curvature}b, and \ref{fig:fixed-frequency-BLDS}, while light-induced changes in the out-of-phase force causes the dissipation phenomena apparent in Figures~\ref{fig:Dissipation-curvature}b, \ref{fig:dissipation-recovery-intensity}, \ref{fig:dissipation-recovery-temperature}, and \ref{dissipation-tip-pulse}.
Above we concluded from the high-frequency  data in Figures~\ref{fig:BLDS-roll-off}a,b and Figure~\ref{fig:fixed-frequency-BLDS} that the light-dependence of both the dissipation and the BLDS response could be attributed to changes in sample resistance $R\st{s}$ (or conductivity $\sigma\st{s}$) rather than sample capacitance $C\st{s}$.
Measuring dissipation thus allowed us to track changes in sample conductivity in real time as a function of light intensity and temperature.
The resulting picture of sample conductivity dynamics was corroborated by monitoring the sample's dielectric spectrum in real time at selected frequencies.

We found that the observed sample conductivity in FAMACs
\begin{enumerate}
\item was substrate dependent, and 
  \begin{enumerate}
  \item was comparatively low in the dark and \emph{dependent} on light intensity over \ce{TiO2} and \ce{SnO2} \label{Obs:low-sigma-dark} but
  \item was comparatively high in the dark and \emph{independent} of light intensity over \ce{ITO} and \ce{NiO}. \label{Obs:high-sigma-dark}
  \end{enumerate}
\end{enumerate}
The light-dependent conductivity of FAMACs grown over \ce{TiO2} and \ce{SnO2} was studied in detail. We found that the conductivity in these samples
\begin{enumerate}
\setcounter{enumi}{1}
\item increased rapidly ($<$ $\SI{0.1}{\second}$) when light was applied; \label{Obs:inc-promptly-with-light}
\item had a steady-state value which increased with light intensity, with this increase being temperature-independent over \ce{SnO2}; \label{Obs:inc-with-light}
\item retained its light-on value when the light was turned off at room temperature
  \begin{enumerate}
  \item for 10's of seconds over \ce{TiO2} and
  \item for 100's of seconds over \ce{SnO2}; 
  \end{enumerate}
  \label{Obs:persistant}
\item relaxed from a light-on value to a light-off value above room temperature with a rate that 
  \begin{enumerate}
  \item increased with increasing temperature over \ce{TiO2}, with a large activation energy, $E_{\text{a}} = 0.58 \: \mathrm{eV}$, usually associated with vacancy or halide-ion motion but
  \label{Obs:recovery-TiO2}
  \item had no measurable temperature dependence over \ce{SnO2}; and
  \label{Obs:recovery-SnO2}
  \end{enumerate}
  \label{Obs:recovery}
\item relaxed from its light-on to its light-off value essentially instantaneously at low temperature. \label{Obs:recovery-LT}
\end{enumerate}
We wish to explain these findings microscopically.
The first step in doing so is to consider the source of the observed conductivity.

\subsection{Conductivity sources}

There are two obvious contributions to sample conductivity $\sigma\st{s}$ to consider --- electronic and ionic.
Light-dependent \emph{electronic} conductivity is expected in a semiconductor like a lead-halide perovskite in which light absorption in the bulk creates free electrons and holes.
If the conductivity is dominated by electronic conductivity then one would expect the conductivity to increase rapidly under illumination and be intensity-dependent, consistent with Observation~\ref{Obs:low-sigma-dark}. 
Further experiments show that the fast time constant is essentially linear in light intensity (Figures~\ref{TiO-time-constant} and \ref{SnO-time-constant}). 
When the light is turned off, however, the electronic conductivity should decay to its light-off value on the timescale of the carrier lifetime --- nanoseconds to microseconds in lead-halide perovskites \cite{Wehrenfennig2014mar,deQuilettes2015may,Hutter2015jul,deQuilettes2016jul,Chen2016aug,Stranks2017jul}.
Instead we find that the light-induced conductivity over \ce{TiO2} and \ce{SnO2} \emph{persisted} for 10's to 100's of seconds when the light was turned off.

The observed sample conductivity $\sigma\st{s}$ could alternatively be dominated by \emph{ionic} conductivity.
Perovskites are expected to have a high concentration of charged vacancies \cite{Yin2014feb,Walsh2015feb,Shi2015mara}; the vacancy concentration depends on the electron Fermi level and on the chemical potential (\latin{i.e.} the concentration) of the chemical species present during film growth \cite{Yin2014feb,Shi2015mara}.
Prior studies have demonstrated that the electron Fermi level of the perovskite can moreover be altered by changing the work function of the substrate \cite{Miller2014oct,Olthof2016sep,Ou2017jan,Olthof2017jan}, with recent work demonstrating that the substrate can change the stoichiometry of the perovskite film as well \cite{Dou2017sep}.
Based on the observations of Refs.~\citenum{Miller2014oct}, \citenum{Olthof2017jan}, \citenum{Olthof2016sep}, and \citenum{Ou2017jan}, and  we would expect a high halide-vacancy concentration in a perovskite grown over a hole-injecting substrate like \ce{NiO} or \ce{ITO}, in agreement with the observed trends in the light-off conductivity, Observation~\ref{Obs:high-sigma-dark}.
If the sample conductivity is dominated by \emph{ionic} conductivity, however, we would not expect the ionic conductivity to be linearly proportional to light intensity and independent of temperature (Observations~\ref{Obs:inc-with-light}), nor would we expect ionic conductivity to retain a memory of the light intensity for 10's to 100's of seconds in the dark (Observations~\ref{Obs:inc-with-light} and \ref{Obs:persistant}).

In summary, the observed conductivity has attributes of both electronic and ionic conductivity.  
Tirmzi and coworkers observed a similarly puzzling long-lived photo-induced conductivity in their related prior dissipation-microscopy experiments on \ce{CsPbBr3}.
They posited that photo-induced electrons and holes were being captured by charged vacancies existing in the film \cite{Tirmzi2017jan}:
\begin{linenomath}
\begin{subequations}
\begin{align}
\text{V}_{\text{Br}}^{\bullet} 
  + \text{e}^{\prime} 
  & \ce{<=>}
  \left( \text{V}_{\text{Br}}^{\bullet}
  \cdots \text{e}^{\prime} \right) \\ 
\text{Br}_{\text{i}}^{\prime} 
  + \text{h}^{\bullet} 
  & \ce{<=>}
  \left( \text{Br}_{\text{i}}^{\prime}
  \cdots \text{h}^{\bullet} \right).
\end{align}
\label{Eq:Tirmzi-hypotheses}
\end{subequations}
\end{linenomath}
The idea of a weakly-trapped electron and hole, $\left( \text{V}_{\text{Br}}^{\bullet} \cdots \text{e}^{\prime} \right)$ and $\left( \text{Br}_{\text{i}}^{\prime} \cdots \text{h}^{\bullet} \right)$ respectively, was proposed as a way to simultaneously account for the conductivity's light dependence, memory, and large activation energy.  
The Eq.~\ref{Eq:Tirmzi-hypotheses} proposal required the $\left( \text{V}_{\text{Br}}^{\bullet} \cdots \text{e}^{\prime} \right)$ and $\left( \text{Br}_{\text{i}}^{\prime} \cdots \text{h}^{\bullet} \right)$ species to dominate the conductivity, which the Ref.~\citenum{Tirmzi2017jan} authors noted was seemingly at odds with the idea of a weakly-trapped electron and hole.
The notion that $\text{V}_{\text{Br}}^{\bullet}$ and $\text{e}^{\prime}$ (or $\text{Br}_{\text{i}}^{\prime}$ and $\text{h}^{\bullet}$) diffuse together as a unit is the central idea underlying the concept of \emph{ambipolar conductivity}, although the authors of Ref.~\citenum{Tirmzi2017jan} did not employ this term.
We will consider ambipolar conductivity in more detail shortly.    

The hypothesis that we are observing ambipolar conductivity resolves some but not all of our puzzling conductivity observations.
We need another key new idea.
Since the work of Tirmzi \latin{et al.}, Kim, Maier, and coworkers \cite{Kim2018mar} have used multiple physical measurements to demonstrate that light induces a large enhancement in the \emph{ionic} conductivity of methylammonium lead iodide.
To explain this observation they proposed a reaction of photo-induced holes with neutral iodine atoms in the lattice that generates neutral interstitial iodines and charged, mobile iodine vacancies:
\begin{linenomath}
\begin{align}
\text{I}_{\st{I}}^{\text{x}} 
+ \text{h}^{\bullet}
& \ce{<=>} 
\text{I}_{\st{i}}^{\text{x}} 
+ \text{V}_{\st{I}}^{\bullet}.
\label{Eq:Kim-Maier}
\end{align}
\end{linenomath}
In their view, the application of light increases the concentration of holes, $[\text{h}^{\bullet}]$, which shifts the Eq.~\ref{Eq:Kim-Maier} equilibrium to the right;
this shift increases the concentration of $\text{V}_{\st{I}}^{\bullet}$ which in turn raises the ionic conductivity.
That the halide-vacancy concentration depends on $[\text{h}^{\bullet}]$ is expected, given the dependence of defect concentration on electron Fermi level \cite{Yin2014feb,Shi2015mara,Senocrate2018aug}.
The significance of the Kim \latin{et al.} data is that it experimentally demonstrates the existence of light-induced changes in ionic conductivity and quantifies the size of the effect.
For our purposes, the Eq.~\ref{Eq:Kim-Maier} observation provides a better starting point for understanding our observations than does the Eq.~\ref{Eq:Tirmzi-hypotheses} conjecture.
To describe our further observations it is helpful to augment Eq.~\ref{Eq:Kim-Maier} to include both the holes and electrons created by light absorption:
\begin{linenomath}
\begin{align}
\text{I}_{\st{I}}^{\text{x}}
\xrightarrow{\: \: h \nu \: \:}
\text{I}_{\st{I}}^{\text{x}} 
+ \text{h}^{\bullet}
+ \text{e}^{\prime}
& \ce{<=>} 
\text{I}_{\st{i}}^{\text{x}} 
+ \text{V}_{\st{I}}^{\bullet}
+ \text{e}^{\prime}.
\label{Eq:Kim-Maier-modified}
\end{align}
\end{linenomath}
Equation~\ref{Eq:Kim-Maier-modified} indicates the presence, after illumination, of cationic vacancies and charge-compensating electrons, both of which are mobile.
We should therefore formulate the sample's dielectric response in terms of its \emph{ambipolar} conductivity \cite{Barsoum2003}.

\subsection{Ambipolar conductivity}
\label{Sec:ambipolar-conductivity}

The relevance of ambipolar conductivity to understanding light-dependent phenomena in mixed ionic-electronic conductors like metal halide perovskites is just becoming apparent \cite{Kerner2017dec}.
In our prior scanned-probe study of \ce{CsPbBr3} we modeled the sample as a resistor $R\st{s}$ and capacitor $C\st{s}$ connected in parallel. 
The quantitative response of an ambipolar sample in an electric force microscope experiment has not, to our knowledge, been considered before \cite{Morozovska2012jan}.
In order to ascertain the dependence of measured dissipation and frequency shift on the sample's electronic and ionic conductivity, in this section we apply a more physically accurate transmission-line model of the sample's dielectric response \cite{Jamnik1999apr}.

The starting point for modeling the response of a charged cantilever to a conductive sample is the transfer function in Eq.~\ref{eq:H-full}, which may be simplified to read
\begin{equation}
\hat{H}(\omega) 
= \frac{1}
  {1 +j \omega C\st{tip} Z\st{s}(\omega)} 
\label{eq:H-Z}
\end{equation}
with $Z\st{s}(\omega)$ the sample impedance.
The impedance of a mixed ionic-electronic conductor was first derived in detail for various electrode models by McDonald \cite{Macdonald1973jun} but the derivation ignored space-charge regions near the contacts.
A more tractable and generalizable transmission-line treatment of a mixed ionic-electronic conductor was introduced by Jamnik and Maier \cite{Jamnik1999apr}.
Their approach has since been applied to calculate the impedance spectra of materials ranging from ion-conducting ceramics \cite{Lai2005nov,Ciucci2009dec} to lead-halide perovskite photovoltaics \cite{Bertoluzzi2014may,Bisquert2014aug}.
Let us use the impedance formula given in Ref.~\citenum{Jamnik1999apr} (correctly written as Eq.~61 in Ref.~\citenum{Lai2005nov}) to calculate an approximate $Z\st{s}(\omega)$ for our sample.

In the Ref.~\citenum{Jamnik1999apr} model, the sample is assumed to contain two mobile charged carriers, where the first species is ionic (charge $z\st{ion}= 1$, concentration $c\st{ion}$, conductivity $\sigma\st{ion}$) and the second species is electronic (charge $z\st{eon} = 1$, concentration $c\st{eon}$, conductivity $\sigma\st{eon}$).
The associated ionic and electronic resistance is given by $R\st{ion} = L \big/ \sigma\st{ion} A$ and $R\st{eon} = L \big/ \sigma\st{eon} A$, respectively, with $L$ the sample thickness and $A$ the sample cross-sectional area.
Two other variables arise naturally in the transmission-line treatment.
The first is the chemical capacitance \cite{Jamnik1999apr},
\begin{equation}
C\st{chem} 
  = \frac{q^{2}}{k\st{b}T}
  \Bigg(
      \frac{1}{z\st{ion}^{2} c\st{ion}^{\vphantom{2}}} 
    + \frac{1}{z\st{eon}^{2} c\st{eon}^{\vphantom{2}}}
  \Bigg)^{-1} 
  A L 
\label{eq:C-chem}
\end{equation}
with $q$ the electronic unit of charge, $k\st{b}$ Boltzmann's constant, and $T$ temperature.
It is reasonable to assume that $c\st{ion} \gg c\st{eon}$ \cite{Shi2015jan,Bi2014sep,deQuilettes2015may,Walsh2015feb}; in this limit, $C\st{chem} \approx  q^2 A L \, c\st{eon} \big/ k\st{b} T$, and the chemical capacitance is determined by the concentration of the electronic carriers alone.
The second central variable is the ambipolar diffusion constant, defined as
\begin{equation}
D\st{a}
  = \frac{k\st{b}T}{q^{2}}
    \frac{\sigma\st{ion}\sigma\st{eon}}  
         {(\sigma\st{ion}+\sigma\st{eon})} 
  \Bigg\{
       \frac{1}{z\st{ion}^{2} c\st{ion}^{\vphantom{2}}} 
     + \frac{1}{z\st{eon}^{2} c\st{eon}^{\vphantom{2}}}
   \Bigg\},
\end{equation}
which simplifies to
\begin{equation}
D\st{a} = 
  \frac{L^{2}}
    {(R\st{ion} + R\st{eon}) \, C\st{chem}}.
\label{eq:Ambi-diffusion}
\end{equation}
In Ref.~\citenum{Jamnik1999apr} the electrodes are assumed to be symmetric and described by a distinct interface impedance for ionic and electrical carriers:
\begin{linenomath}
\begin{subequations} 
\begin{align}
Z\st{ion}^{\bot} 
  & = \frac{R\st{ion}^{\bot} \vphantom{\big|}}
      {1+j\omega R\st{ion}^{\bot}C\st{ion}^{\bot}} \\
Z\st{eon}^{\bot} 
  & = \frac{R\st{eon}^{\bot} \vphantom{\big|}}
      {1+j\omega R\st{eon}^{\bot}C\st{eon}^{\bot}}
\end{align}
\end{subequations}
\end{linenomath} 
with the subscript indicating the carrier and the superscript $\bot$ indicating that the resistance and capacitance is associated with the sample/electrode interface.
We can rearrange Jamnik and Maier's central impedance result to read
\begin{linenomath}
\begin{multline}
Z\st{s}(\omega) 
 = Z\st{\infty} + (Z\st{0} - Z\st{\infty}) \\ 
\times 
\frac{R\st{ion} + R\st{eon} + 
       2(Z\st{ion}^{\bot} +Z\st{eon}^{\bot})}
      {R\st{ion} + R\st{eon} + 
       2 (Z\st{ion}^{\bot} +Z\st{eon}^{\bot})
        \sqrt{\frac{j \omega \tau}{4}}
        \coth{\sqrt{\frac{j \omega \tau}{4}}}}
\label{eq:Approx-impedance}
\end{multline}
\end{linenomath}
with the low- and high-frequency limiting impedance given by
\begin{linenomath}
\begin{subequations}
\begin{align}
\frac{1}{Z\st{0}} 
  & = \frac{1}{R\st{ion} + 2 Z\st{ion}^{\bot}} 
    + \frac{1}{R\st{eon} + 2 Z\st{eon}^{\bot}} \\
Z\st{\infty} 
  & = \frac{R\st{ion}   R\st{eon}}
           {R\st{ion} + R\st{eon}} 
  + 2 \frac{Z\st{ion}^{\bot}   Z\st{eon}^{\bot}}
           {Z\st{ion}^{\bot} + Z\st{eon}^{\bot}}
\end{align}
\end{subequations}
\end{linenomath}
and the time constant $\tau$ defined as
\begin{equation}
\tau
  = \frac{L^2}{D\st{a}}
  = (R\st{ion} + R\st{eon}) \, C\st{chem}.
\end{equation}

\begin{figure}[t]
\includegraphics[width=3.25in]{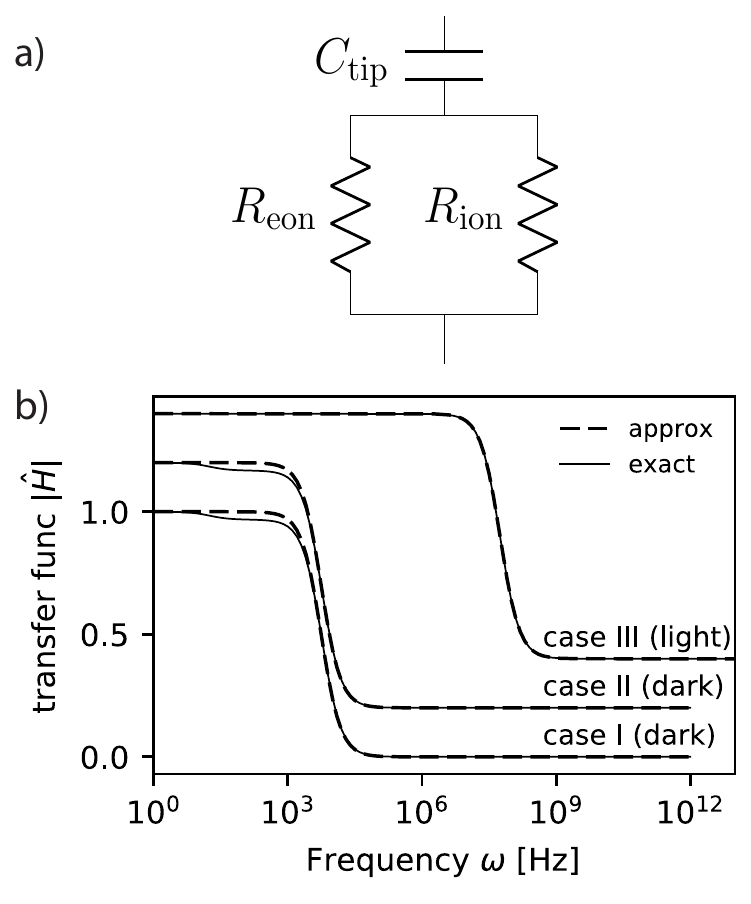}
	\caption{(a) Equivalent circuit representation of the impedance of a mixed ionic-electronic conductor in the high-frequency limit. (b) Approximate (dashed  line) and exact (solid line) transfer function |$\hat{H}$| for the representative case I (dark), case II (dark), and case III (light) sample properties given in Table~\ref{tab:combined}.
	In each case in (b) the transfer function has been offset by $0.2$ for clarity.}
	\label{fig:circuit-impedance-interface}    
\end{figure}

Our sample has a bottom contact consisting of a grounded electrical conductor and a top contact consisting of an electrically biased tip-sample capacitor.
The impedance of the tip-sample capacitor operating in series with the electrically ground sample is already captured in Eq.~\ref{eq:H-Z}.
To capture the impedance of our sample in the transmission-line formalism we assume that the electrodes (1) are ohmic for the electronic carriers, $Z\st{eon}^{\bot} \approx 0$, and (2) are blocking for the ions, $R\st{ion}^{\bot} \rightarrow \infty$ and consequently $Z\st{ion}^{\bot} \approx 1 \big/ j \omega C\st{ion}^{\bot}$.
Under these simplifying assumptions, $Z_{\infty} = R\st{ion} R\st{eon}/(R\st{ion} + R\st{eon})$.
In reality the sample's bottom face is metal-terminated while its top face is vacuum terminated; although the sample is not strictly symmetric, under our electrode assumptions the transmission-line impedance model should nevertheless give accurate guidance on what sample properties our scanned probe measurements are probing.
Substituting for the expressions for $Z_0$ and $Z_{\infty}$ in the ion-blocking limit in Eq.~\ref{eq:Approx-impedance} and simplifying the result we obtain
\begin{equation}
Z\st{s} \approx \frac{R\st{ion}R\st{eon}}{R\st{ion}+ R\st{eon}} \Bigg
\lbrace 1 + \frac{1}{\sqrt{\frac{j\omega}{\omega\st{chem}}} \coth{\sqrt{\frac{j\omega R\st{eon}^{2}}{\omega\st{chem} R\st{ion}^{2}}}} + j\frac{\omega}{\omega\st{ion}}} \Bigg \rbrace
\label{Eq:Z-exact} \\
\end{equation}
with
\begin{equation}
\omega\st{ion} \equiv \left( \frac{1}{2} C\st{ion}^{\bot} \frac{R\st{ion}}{R\st{eon}}(R\st{ion} + R\st{eon}) \right)^{-1}
\label{eq:omega-ion}
\end{equation}
and
\begin{equation}
\omega\st{chem} \equiv \left( \frac{1}{4} C\st{chem} \frac{R\st{ion}^{2}}{R\st{eon}^{2}}(R\st{ion} + R\st{eon}) \right)^{-1}.
\label{eq:omega-chem}
\end{equation}
As long as $\omega \gg \omega\st{chem}$ or $\omega \gg \omega\st{ion}$, the sample impedance $Z\st{s}(\omega)$ will be operating in the high frequency limit where $Z\st{s}$ $\approx$ $Z\st{\infty}$. 
The transfer function describing the tip-sample interaction in this high-frequency limit can be approximated as (Figure~\ref{fig:circuit-impedance-interface}a)
\begin{linenomath}
\begin{subequations}
\label{Eq:H-approx}
\begin{align}
\hat{H}(\omega)
 & \approx
 \frac{1}{1 + j \, \omega \, \tau\st{tip}} 
    \label{Eq:H-approx-H} \\
\intertext{with}
\tau\st{tip}
  & = \omega\st{tip}^{-1} 
  = \frac{L}{A} 
   \frac{C\st{tip}}
        {\sigma\st{ion} + \sigma\st{eon}}.
\label{eq:omega-tip}   
\end{align}
\end{subequations}
\end{linenomath} 
Somewhat surprisingly, the rolloff of $\hat{H}(\omega)$ depends not on the ambipolar conductivity but on the \emph{total} conductivity, $\sigma\st{ion} + \sigma\st{eon}$.
\begin{table*}
\renewcommand{\arraystretch}{1.2}
\begin{tabular}{c c c c c c c c c}
\hline
 &
 &
 & \multicolumn{2}{c}{case I (dark)} 
 & \multicolumn{2}{c}{case II (dark)}
 & \multicolumn{2}{c}{case III (light)}\\
\hline
 & quantity 
 & unit 
 & value 
 & Ref 
 & value 
 & Ref 
 & value 
 & Ref\\
\hline
A. 
 & $c\st{eon}$ 
 & $\SI{}{\per\centi\meter\cubed}$ 
 & $\SI{5d9}{}$
 & \citenum{Shi2015jan}
 & $\SI{6d14}{}$
 & \citenum{Bi2014sep} 
 & $\SI{2d15}{}$ 
 & \citenum{deQuilettes2015may} \\

 & $c\st{ion}$ 
 & $\SI{}{\per\cubed\centi\meter\cubed}$ 
 & $\SI{2d20}{}$ 
 & \citenum{Walsh2015feb} 
 & $\SI{2d20}{}$ 
 & \citenum{Walsh2015feb} 
 & $\SI{2d20}{}$ 
 & \\
 
 & $\sigma\st{eon}$ 
 & $\SI{}{\siemens\per\centi\meter}$ 
 & $\SI{5d-9}{}$ 
 & \citenum{Kim2018mar} 
 & $\SI{5d-9}{}$ 
 & \citenum{Kim2018mar} 
 & $\SI{5d-4}{}$ 
 & \citenum{Kim2018mar}\\
 
 & $\sigma\st{ion}$ 
 & $\SI{}{\siemens\per\centi\meter}$ 
 & $\SI{5d-8}{}$ 
 & \citenum{Kim2018mar} 
 & $\SI{5d-8}{}$ 
 & \citenum{Kim2018mar} 
 & $\SI{5d-5}{}$ 
 & \citenum{Kim2018mar}\\

\hline
   
B.
    & $C\st{chem}$ 
    & $\SI{}{\farad}$ 
    & $\SI{1.6d-21}{}$
    &
    & $\SI{1.9d-16}{}$
    &
    & $\SI{6.2d-16}{}$ 
    & \\

\hline

C.  
    & $\omega\st{chem}$
    & $\SI{}{\hertz}$
    & $\SI{1.0d3}{}$
    &
    & $\SI{1.0d3}{}$
    &
    & $\SI{3.0d6}{}$
    &\\
    
    & $\omega\st{ion}$
    & $\SI{}{\hertz}$
    & $\SI{9.2d0}{}$
    &
    & $\SI{1.2d8}{}$
    &
    & $\SI{9.2d4}{}$
    &\\
    
    & $\omega\st{tip}$
    & $\SI{}{\hertz}$
    & $\SI{5.6d3}{}$
    &
    & $\SI{5.6d3}{}$
    &
    & $\SI{5.6d7}{}$
    &\\
         
         \hline
    \end{tabular}
\caption{ (A) Literature values for sample properties needed to calculate (B) and (C).}
\label{tab:combined}
\renewcommand{\arraystretch}{1.0}
\end{table*}

We can check the validity of the approximate Eq.~\ref{Eq:H-approx} transfer function by comparing with $\hat{H}(\omega)$ calculated using the full impedance expression of Eq.~\ref{Eq:Z-exact}.
To calculate the transfer function requires knowledge of $c\st{ion}$, $c\st{eon}$, $\sigma\st{ion}$, $\sigma\st{eon}$, $C\st{ion}^{\bot}$, and $C\st{tip}$.
To our knowledge, no single study provides values for all these quantities for FAMACs.
We therefore turn to the MAPI literature for order-of-magnitude estimates of these quantities; see Table~\ref{tab:combined}A.
The dark $c\st{eon}$ estimates vary from $\SI{5d9}{\per\centi\meter\cubed}$ (case I (dark)) to $\SI{6d14}{\per\centi\meter\cubed}$ (case II (dark)).
For $c\st{ion}$ under illumination (case III) we expect the value to be similar or higher than the corresponding dark value.
We calculated $R\st{eon}$ and $R\st{ion}$ from the Ref.~\citenum{Kim2018mar} conductivities taking $L = \SI{700}{\nano\meter}$, the film thickness, and $A = \SI{7d-14}{\meter\squared}$, our estimate of the cantilever-tip area.
In Table~\ref{tab:combined}A we have linearly scaled the conductivities observed in Ref.~\citenum{Kim2018mar} to account for the higher light intensities used in our measurements.
The value of $C\st{chem}$ (Table~\ref{tab:combined}B) was calculated using Eq.~\ref{eq:C-chem} and the values for $c\st{ion}$ and $c\st{eon}$ given in Table~\ref{tab:combined}A.
We take $C\st{ion}^{\bot} = \epsilon\st{s}\epsilon\st{0}A/\lambda\st{D} \approx \SI{1e-14}{\farad}$ where $\epsilon\st{s} = 26$ \cite{Frost2014may} is the static dielectric constant and $\lambda\st{D} = \SI{1.5}{\nano\meter}$ \cite{Richardson2016apr} is the Debye length.
We use $C\st{tip} = \SI{1d-16}{\farad}$, a reasonable upper-bound number taking in account the experimental tip-sample separation \cite{Gomila2008jul,Dwyer2018jul}.
Using Table~\ref{tab:combined}A-B values and the above estimates for $C\st{ion}^{\bot}$ and $C\st{tip}$ we obtain the  frequencies $\omega\st{tip}$, $\omega\st{ion}$, and $\omega\st{chem}$ given in Table~\ref{tab:combined}C.

We plot the resulting approximate and exact transfer function $\hat{H}(\omega)$ for two dark conditions and one light condition in Figure~\ref{fig:circuit-impedance-interface}b.
This exercise confirms that $Z\st{s} \approx Z\st{\infty}$ is indeed a valid approximation.
The effect of $\omega\st{ion}$ on the transfer function $\hat{H}(\omega)$ is only significant when $\omega\st{ion}$ or $\omega\st{chem}$ are within an order of magnitude of $\omega\st{tip}$.
A slight breakdown of the Eq.~\ref{Eq:H-approx} approximation can be seen in the Figure~\ref{fig:circuit-impedance-interface}b transfer-function plots for case I (dark) and II (dark) at low frequency. 
In most scenarios this breakdown is unlikely to occur because $C\st{ion}^{\bot} \gg C\st{tip}$ and in this limit $\omega\st{tip}$ $\gg$ $\omega\st{ion}$.

\subsection{Explaining the  
  conductivity findings}

Now that we have established that the measurements in this manuscript probe \textit{total conductivity}, we will look at how the concentrations of $\text{h}^{\bullet}$ and $\text{V}_{\text{I}}^{\bullet}$ and the Eq.~\ref{Eq:Kim-Maier-modified} scheme can be used to rationalize differences in the conductivity and conductivity relaxation between different substrates.

We would expect the concentration of holes in the dark $[\text{h}^{\bullet}]_{\text{dark}}$ (and therefore $[\text{V}_{\text{I}}^{\bullet}]_{\text{dark}}$) to be high over the \ce{ITO} and \ce{NiO} and low over \ce{TiO2} and \ce{SnO2} \cite{Miller2014oct,Olthof2016sep,Ou2017jan,Olthof2017jan}.
Observations~\ref{Obs:low-sigma-dark}, \ref{Obs:high-sigma-dark}, and \ref{Obs:inc-with-light} follow from Eq.~\ref{Eq:Kim-Maier} and the assumption that $[\text{h}^{\bullet}]_{\text{light}} \ll [\text{h}^{\bullet}]_{\text{dark}}$ over \ce{ITO} and \ce{NiO} while $[\text{h}^{\bullet}]_{\text{light}} \gg [\text{h}^{\bullet}]_{\text{dark}}$ over \ce{TiO2} and \ce{SnO2}.
A change in the sample conductivity due to the substrate is indirectly implied in the results of Refs.~\citenum{Miller2014oct}, \citenum{Olthof2017jan}, \citenum{Olthof2016sep}, and \citenum{Ou2017jan} where the work function of the perovskite surface was shown to change as a function of substrate work function.
Our data likewise shows a substrate effect, only here we probe the conductivity directly.
The high absorption coefficient of the perovskite means that electrons and holes are primarily generated in the top $\sim\! \SI{200}{\nano\meter}$ of our $\SI{700}{\nano\meter}$-thick films.
Possible processes that may exist and can directly or indirectly change the material and therefore the total conductivity include substrate induced strain effects,\citep{Zhao2017Nov,Tsai2018apr} substrate dependent sample microstructure and stoichiometery,\citep{Dou2017sep} and heterogeneous doping.\citep{Maier1986jan}
Our current results are largely inconsistent with heterogeneous doping effects.
Substrate effects through heterogeneous doping are going to be limited to a thin layer near the perovskite-substrate interface and are more prominent when the substrates is mesoporous.
In this layer, the concentration of both electronic and ionic charges is determined by the substrate perovskite interaction.\citep{Maier1986jan}
This is inconsistent with the Figure~\ref{dissipation-tip-pulse} results and the thickness of the films ($\SI{700}{\nano\meter}$) used in our measurements.

Under illumination the concentration of $\text{h}^{\bullet}$ is high; the forward reaction in Eq.~\ref{Eq:Kim-Maier-modified} proceeds rapidly, creating charged vacancies and free electrons resulting in the promptly appearing light-dependent conductivity of Observations~\ref{Obs:inc-promptly-with-light} and \ref{Obs:inc-with-light}.
The relative similarity of the dependence of the conductivity on time, light, and temperature over \ce{SnO2} and \ce{TiO2} suggests to us that the light-on conductivity in \ce{TiO2} is also likely dominated by electronic conductivity. 
According to Eq.~\ref{eq:omega-tip}, for the observed \emph{total} conductivity to return to its light-off value, both $\sigma\st{ion}$ and $\sigma\st{eon}$ need to return to their dark values.
Observation~\ref{Obs:persistant} is explained by the back reaction in Eq.~\ref{Eq:Kim-Maier-modified} having a high activation energy and proceeding slowly.
The differences in the timescale of the conductivity relaxation over \ce{TiO2} and \ce{SnO2}, Observations~\ref{Obs:recovery-TiO2} and \ref{Obs:recovery-SnO2}, requires a difference in this activation energy or in the mobility of ions in the FAMACs grown on the two substrates.
Christians and coworkers have reported a differences in the distribution of ions in aged devices incorporating \ce{TiO2}/FAMACs and \ce{SnO2}/FAMACs interfaces, as quantified by time-of-flight secondary ion mass spectrometry \cite{Christians2018jan}; these differences are consistent with the slower relaxation seen here over \ce{SnO2}.

The fast conductivity relaxation seen at low temperature, Observation~\ref{Obs:recovery-LT}, seems \latin{prima facia} at odds with the slow and activated recovery seen at room temperature, Observations~\ref{Obs:persistant} and \ref{Obs:recovery}.
We should consider, however, that once generated, the iodine vacancy, $\text{V}_{\text{I}}^{\bullet}$, and the neutral iodine interstitial defect, $\text{I}_{\text{i}}^{\text{x}}$, are expected to diffuse away from each other (to maximize entropy).
In mixed-halide perovskites light has been shown to promote halide segregation and in these systems the rate of segregation depends on the light intensity \cite{Hoke2015jan,Draguta2017dec}. 
One might therefore expect the Eq.~\ref{Eq:Kim-Maier-modified} back reaction underlying Observations~\ref{Obs:persistant} and \ref{Obs:recovery} to be diffusion limited; in this limit the activation energy of the back reaction is the $E_{\text{a}}$ governing $\text{V}_{\text{I}}^{\bullet}$ and $\text{I}_{\text{i}}^{\text{x}}$ diffusion.
The $E_{\text{a}}$ we observe over \ce{TiO2} is consistent with the activation energy measured for halide-vacancy motion in lead-halide perovskites \cite{Mizusaki1983nov,Hoke2015jan,Yang2015jun}.
The activation energy observed is the activation energy for the total conductivity relaxation.
This will in turn depend on the concentration electronic and ionic species, but also on their mobility.
We note that the light intensity primarily determines the concentration of both ionic and electronic carriers and was kept constant for variable temperature measurements.
While we are not directly probing the activation energy of ionic diffusion (and therefore the ionic mobility), it is the most likely term to change in the small temperature window used in the measurement.
At low temperature we expect the vacancy diffusion to be suppressed and consequently might expect the back reaction to be now fast because the $\text{V}_{\text{I}}^{\bullet}$ and $\text{I}_{\text{i}}^{\text{x}}$ species generated by the forward Eq.~\ref{Eq:Kim-Maier-modified} reaction remain in close proximity.
This prediction is indeed consistent with Observation~\ref{Obs:recovery-LT}.

Subsequent reactions are also possible.
Based on Minns \latin{et al.}'s \cite{Minns2017may} X-ray and neutron diffraction studies of \ce{(CH3NH3)PbI3}, for example, we expect the iodine interstitials to form stable interstitial \ce{I2} moieties.
The concentration of these \ce{I2} moieties (and the coupled vacancy concentration) can be decreased by lowering the temperature.
Additionally, theory predicts the iodine interstitial to be a hole trap, $\text{I}_{\text{i}}^{\text{x}} + \text{h}^{\bullet} \ce{<=>} \text{I}_{\text{i}}^{\bullet}$ \cite{Li2017jun}.
Such reactions and the decreased concentration of \ce{I2} moieties, if present, might likewise explain the significant differences in recovery seen over \ce{TiO2} and \ce{SnO2}, Observations~\ref{Obs:recovery-TiO2} and \ref{Obs:recovery-SnO2}.


\section{Conclusions}
Here we have used measurements of sample-induced dissipation and sample dielectric spectra, backed by a rigorous theory of the cantilever-sample interaction \cite{Dwyer2017jun,Tirmzi2017jan,Dwyer2018jul}, to carry out time-resolved studies of photo-induced changes in the total conductivity of a mixed-species lead-halide perovskite semiconductor thin film prepared on a range of substrates.
Comparison of low temperature and room temperature data and a transmission-line model analysis of mixed ionic-electronic conductivity reveals that the observed photo-induced changes in cantilever frequency and dissipation report on changes in total sample conductivity, $\sigma\st{ion} + \sigma\st{eon}$.
This insight establishes scanning-probe broadband local dielelectric spectroscopy measurements as a method for quantifying local photo-conductivity in semiconductors and other photovoltaic materials.

In the FAMACs samples studied here, light-induced changes in total conductivity relaxed on a time scale of $10$'s to $100$'s of seconds, with an activation energy of \SI{0.58}{\eV} over \ce{TiO2}; such a large activation energy is generally attributed to ion/vacancy motion \cite{Mizusaki1983nov,Hoke2015jan,Yang2015jun}.
We rationalized these findings using the idea of light-induced vacancies recently proposed by Kim \latin{et al.}\cite{Kim2018mar}
In addition to the seemingly puzzling light-induced conductivity behavior explored here, light-induced creation of vacancies may also explain other light-induced anomalous behavior seen in lead halide perovskites including memory effects \cite{Belisle2017jan}.


\bigskip

\noindent {\large \textbf{SUPPORTING INFORMATION AVAILABLE}} 
The Supporting Information contains:
Experimental details regarding scanning probe microscopy; experimental details for Fig.~\ref{fig:Dissipation-curvature}; spatial variation the BLDS response; representative AFM images; steady state and transient surface potential; fit detail for Fig.~\ref{fig:dissipation-recovery-temperature}; effect of below band gap illumination on dissipation; $\tau\st{\Gamma_{\mathrm{s}}}$ for \ce{SnO2}-substrate sample.


\bigskip

\noindent {\large \textbf{AUTHOR INFORMATION}}

\bigskip


\medskip

\noindent \textbf{Corresponding Author}

\medskip

\noindent ${}^{*}$E-mail: \href{mailto:jam99@cornell.edu}{jam99@cornell.edu}

\noindent Faculty webpage: \\
\href{http://chemistry.cornell.edu/john-marohn}{http://chemistry.cornell.edu/john-marohn}

\noindent Research group webpage: \\
\href{http://marohn.chem.cornell.edu/}{http://marohn.chem.cornell.edu/}

\medskip


\medskip

\medskip

\noindent \textbf{Notes}

\medskip

\noindent The authors declare no competing financial interest.

\medskip

\begin{acknowledgement}
A.M.T, R.P.D, and J.A.M acknowledge the financial support of the U.S. National Science Foundation (Grant
DMR-1709879).
J.A.C. was supported by the Department of Energy (DOE) Office of Energy Efficiency and Renewable Energy (EERE) Postdoctoral Research Award under the EERE Solar Energy Technologies Office administered by the Oak Ridge Institute for Science and Education (ORISE) for the DOE under DOE contract number DE-SC00014664.
\end{acknowledgement}
\providecommand{\latin}[1]{#1}
\makeatletter
\providecommand{\doi}
  {\begingroup\let\do\@makeother\dospecials
  \catcode`\{=1 \catcode`\}=2 \doi@aux}
\providecommand{\doi@aux}[1]{\endgroup\texttt{#1}}
\makeatother
\providecommand*\mcitethebibliography{\thebibliography}
\csname @ifundefined\endcsname{endmcitethebibliography}
  {\let\endmcitethebibliography\endthebibliography}{}

\begin{suppinfo}
\section{Experimental Section}
\subsection{Broadband local dielectric spectroscopy details}

To record broadband local dielectric spectroscopy (BLDS) spectra \citep{Labardi2016maya} we applied the following a time-dependent voltage to the cantilever tip:
\begin{equation}
V\st{m}(t) 
	= V\st{pp} 
	\left(  \frac{1}{2} + \frac{1}{2} \cos(2 \pi f\st{am} t) \right)
	\: \cos(2\pi f\st{m} t).
\label{eq:tip-modulation-BLDS}
\end{equation}
The time-dependent voltage in Equation~\ref{eq:tip-modulation-BLDS} consists of a frequency-$f\st{m}$ cosine wave  multiplied by a frequency-$f\st{am}$ amplitude-modulation function.
In the experiments reported in the manuscript, $f\st{m} = \SI{200}{\hertz}$ to $\SI{1.5}{\mega\hertz}$, $f\st{am} = \SI{45}{\hertz}$, and the amplitude was set to $V\st{pp} = \SI{6}{\V}$.
The time-dependent voltage in Equation~\ref{eq:tip-modulation-BLDS} was generated using a digital signal generator (Keysight 33600).
In implementing the BLDS experiment using the time-dependent voltage in Equation~\ref{eq:tip-modulation-BLDS} we are replacing the simple ON/OFF amplitude-modulating function of Ref.~\citenum{Tirmzi2017jana} with a smoother, sinusoidal amplitude-modulating function.
The time-dependent voltage of Equation~\ref{eq:tip-modulation-BLDS} gives rise to a time-dependent cantilever frequency.
According to the theory presented in Ref.~\citenum{Tirmzi2017jana}, the Fourier component of the cantilever frequency at $f\st{am}$, $\Delta f\st{BLDS}$, reports on the tip-sample transfer function $H$ evaluated at $f\st{m}$, $f\st{c} - f\st{m}$, and $f\st{c} + f\st{m}$ (Equation~\ref{eq:Deltaf-BLDS}).
To obtain the data in Figure~\ref{fig:BLDS-roll-off} and \ref{BLDS-variable-temp}, the cantilever frequency shift was measured in real time using a phase-locked loop (PLL; RHK Technology, model PLLPro2 Universal AFM controller), the output of which was fed into a lock-in amplifier (LIA; Stanford Research Systems, model $830$).
The LIA time constant and filter bandwidth were $\SI{300}{\ms}$ and $\SI{6}{\dB}/\mathrm{oct}$, respectively. 
At each stepped value of $f\st{m}$, a wait time of $\SI{1500}{\ms}$ was employed, after which frequency-shift data were recorded for an integer number of frequency cycles corresponding to $\approx \SI{2}{\sec}$ of data acquisition at each $f\st{m}$.

The $\Delta f\st{BLDS}$ frequency-shift signal was obtained from the LIA outputs as follows.
From the (real) in-phase and out-of-phase voltage signals $V_X$ and $V_Y$, respectively, a single (complex signal) in hertz was calculated using the formula
\begin{equation}
Z\st{Hz} = \left (V_x + j \, V_y  \right) \frac{S}{10} \times \sqrt{2} \times 20 \, \frac{\text{Hz}}{\text{V}}
\end{equation}
where the first factor converts from volts-output to volts-input using the sensitivity factor $S$ ($S = 1$ here);
the second factor $\sqrt{2}$ converts the root-mean-square voltage output by the LIA into a zero-to-peak voltage;
and the third factor converts from volts to hertz using the PLL's sensitivity of \SI{20}{\hertz\per\volt}.
From $Z\st{Hz}$ we calculate 
\begin{equation}
\alpha = \frac{4 \lvert Z\st{Hz}\rvert}{V\st{pp}^2}
 = \frac{8 \sqrt{2} \, S \sqrt{V_x^2 + V_y^2}}{V\st{pp}^2}.
\label{eq:alpha-calculation}
\end{equation}
In our experiments $V\st{pp}$ is the estimated voltage at the tip which can be different from the output voltage at the generator due to differences in the output impedance of the signal generator and the impedance of the wiring leading up to the to the cantilever tip.

For fixed-frequency dielectric response curves, a similar method for acquisition was used except that the modulation frequency was not swept and the response was monitored at a single frequency.
The minimum value (which was the dark value) was used as a baseline to calculate the change $\Delta\alpha$ in $\alpha$.

\section{Scanning probe microscopy details}

All experiments were performed under vacuum (\SI{5e-6}{\milli\bar}) in a custom-built scanning Kelvin probe microscope described in detail elsewhere \cite{Luria2011auga, Dwyer2017juna}.

The cantilever was a MikroMasch HQ:NSC18/Pt conductive probe.
The resonance frequency and quality factor were obtained from ringdown measurements and found to be $\omega\st{c}/2\pi = f\st{c} = \SI{70.350}{\kilo\Hz}$ and $Q = \num{24000}$ respectively.
Some experiments were performed with a cantilever having $\omega\st{c}/2\pi = f\st{c} = \SI{72.600}{\kilo\Hz}$ and $Q = \num{29000}$.
The manufacturer's specified resonance frequency and spring constant were $f\st{c} = 60$ to $\SI{75}{\kilo\Hz}$ and $k = \SI{3.5}{\N\per\m}$.
In our analysis, we assumed a spring constant of $k = \SI{3.5}{\N\per\m}$.

Cantilever motion was detected using a fiber interferometer operating at \SI{1490}{\nm} (Corning model SMF-28 fiber).
The DC current of the laser diode (QPhotonics laser diode model QFLD1490-1490-5S) was set using a precision current source (ILX Lightwave model LDX-3620) and the current was modulated at radiofrequencies using the input on the laser diode mount to improve stability and lower noise (ILX Lightwave model LDM-4984, temperature controlled with ILX Lightwave model LDT-5910B) \cite{Rugar1989deca,Muller2005auga}.
The interferometer light was detected with a \SI{200}{kHz} bandwidth photodetector (New Focus model 2011) and digitized at \SI{1}{\mega\hertz} (National Instruments model PCI-6259).
The cantilever was driven using a commercial phase locked loop (PLL) cantilever controller (RHK Technology model PLLPro2 Universal AFM controller), with typical PLL loop bandwidth $\SI{500}{\Hz}$ (PLL feedback loop integral gain $I$ $=$ $\SI{2.5}{\per\Hz}$, proportional gain $P = \SI{-10}{degrees\per\Hz}$).
Frequency and amplitude were determined by software demodulation \cite{Yazdanian2008juna,Dwyer2015jana}.

DC voltages were applied to the tip or sample with either a digital output from the PCI-6259 or a Keithley model 2400 source unit.
The sample was illuminated from above with a fiber-coupled $\SI{639}{\nano\meter}$ laser (Thorlabs model LP635-SF8, held at $\SI{25}{\celsius}$ with a Thorlabs model TED200C temperature controller).
The laser current was controlled using the external modulation input of the laser's current controller (Thorlabs model LDC202, $\SI{200}{\kilo\Hz}$ bandwidth).
The light was coupled to the sample through a multimode, $\SI{50}{\micro\meter}$ diameter core, $0.22$ NA optical fiber (Thorlabs model FG050LGA).
The light source in the Figure~\ref{fig:dissipation-recovery-intensity}a experiment was a Thorlabs model LP520-SF15 laser.
The light source in the Figure~\ref{Dissipation-below-bandgap} experiment was a Thorlabs model LP980-SF15 laser.
The illumination intensity at the sample was calculated from measured power and the estimated spot size at the sample \cite{Dwyer2017juna}.

\section{Frequency shift \latin{vs.}~voltage and amplitude \latin{vs.}~voltage curves}
\label{sec:curvature-dissipation-calculation}
We followed the methods described in Ref.~\citenum{Tirmzi2017jana} to calculate (1) $\alpha\st{0}$ from cantilever frequency shift ($\Delta f$) \latin{vs}.~applied tip voltage ($V\st{ts}$) data and (2) $\gamma\st{s}$ from cantilever amplitude ($A$) \latin{vs}.~applied tip voltage ($V\st{ts}$) data.
Measurements were performed with a constant excitation force $F\st{ex}$. 

The cantilever frequency shift depends on the applied voltage according to the equation
\begin{align}
\Delta f(V\st{ts}) = \alpha\st{0}(V\st{ts}-\phi)^2 
\label{eq:fq-vs-voltage}
\end{align}
with $\phi$ the sample surface potential and $\alpha\st{0}$ a curvature having units of $\SI{}{\hertz\per\volt\squared}$.
The sample surface potential $\phi$ and curvature  $\alpha\st{0}$ were obtained by measuring the cantilever frequency shift $\Delta f$ \latin{vs.}~the applied tip-sample voltage $V\st{ts}$ and fitting the resulting data to Equation~\ref{eq:fq-vs-voltage}.

With the applied excitation force $F\st{ex}$ held constant, the cantilever's steady-state amplitude $A$ is related to the cantilever dissipation $\Gamma$ as follows:
\begin{equation}
F\st{ex} 
  = A \, \Gamma \omega\st{c},
 \label{eq:const-F}
\end{equation}
with $\omega\st{c}$ the cantilever resonance frequency.
The total dissipation experienced by the cantilever is the sum of an intrinsic dissipation $\Gamma\st{i}$ and a sample-induced dissipation $\Gamma\st{s}$:
\begin{equation}
\Gamma = \Gamma\st{i} + \Gamma\st{s}.
\label{eq:Gamma-total}
\end{equation}
The cantilever's intrinsic dissipation, 
\begin{equation}
\Gamma\st{i} 
  = \frac{k}{\omega\st{c} Q}
  \approx \SI{207}{\pico\N\s\per\m},
  \label{eq:Gamma-i}
\end{equation}
was determined from the measured $\omega\st{c}$ and $Q$ assuming $k = \SI{3.5}{\N\per\m}$.
The sample-induced dissipation is found to depend on the applied tip-sample voltage as follows:
\begin{equation}
\Gamma\st{s}(V\st{ts}) 
  = \gamma\st{s}
  (V\st{ts}-\phi)^2,
  \label{eq:Gamma-s}
\end{equation}
with $\gamma\st{s}$ a voltage-normalized, sample-dependent dissipation constant.
Because the total dissipation $\Gamma$ depends on the tip-sample voltage, according to Equation~\ref{eq:const-F} the amplitude will also depend on the tip-sample voltage.
Inserting Equation~\ref{eq:Gamma-s} and \ref{eq:Gamma-i} into Equation~\ref{eq:Gamma-total}, inserting the result into Equation~\ref{eq:const-F}, and solving for the voltage-dependent cantilever amplitude, we obtain
\begin{equation}
A(V)
  = \frac{A_0 \Gamma\st{i}}
  {\Gamma\st{i} + \gamma\st{s}
    (V\st{ts} - \phi)^2},
  \label{eq:gamma-s-vs-voltage}
\end{equation} 
with $A_0$ the cantilever amplitude when the drive voltage $V\st{ts}$ is set to $\phi$.
The voltage-normalized sample-induced dissipation $\gamma\st{s}$, cantilever peak amplitude $A_0$, and sample surface potential $\phi$ were obtained by measuring the cantilever amplitude $A$ as a function of the tip-sample voltage $V\st{ts}$ and fitting the resulting data to Equation~\ref{eq:gamma-s-vs-voltage}.
To acquire this data, a wait time between voltage points of \SI{500}{\ms} --- greater than 3 times the \SI{145}{\ms} cantilever ring down time --- was used so that the measured amplitude was the steady-state amplitude.

\section{Dissipation-recovery time constant}
\label{sec:Q-recovery}

For the experiments of Figure~\ref*{dissipation-tip-pulse}, the time-dependent cantilever quality factor $Q(t)$ was converted into a sample-induced dissipation $\Gamma\st{s}$ using the equation
\begin{equation}
\Gamma\st{s}(t) = \frac{k}{\omega\st{c}} \left ( \frac{1}{Q(t)} - \frac{1}{Q_{0}} \right),
\label{eq:Gamma-t}
\end{equation}
where $k\st{c} = \SI{3.5}{\N\per\m}$ is the cantilever spring constant, $\omega\st{s} = 2 \pi \times \SI{70.350}{\kilo\Hz}$ is the cantilever resonance frequency, and $Q_{0}$ is the cantilever quality factor with no voltage applied and the cantilever positioned far away from the surface ($Q_0 \sim \num{24000}$, typically).
The sample-induced dissipation depends on the sample conductivity which in turn depends on both the light intensity and the time $t$ since the light was turned off.
The dependence of $\Gamma\st{s}$ on sample conductivity arises through the dependence of $\Gamma\st{s}$ on the sample-dependent time constant $\tau\st{fast} \approx R\st{s} C\st{tip}$, where $R\st{s}$ is the sample resistance and $C\st{tip}$ is the tip-sample capacitance.
To describe the observed time-dependent dissipation, we allowed the time constant $\tau\st{fast}$ to be time dependent:
\begin{equation}
\Gamma\st{s}(t)
  = \frac{2\Gamma\st{max} 
      \, \omega\st{c} \, \tau\st{fast}(t)}
    {1 + {\big(\omega\st{c} \,
         \tau\st{fast}(t) \big)}^2},
  \label{eq:Gamma-model}
\end{equation}
with $\Gamma\st{max}$ the peak dissipation (occurring when $\tau\st{fast} = 1/\omega\st{c}$).
The sample conductivity $\sigma\st{s}$ is inversely proportional to the sample resistance $R\st{s}$, $\sigma\st{s} \propto 1/R\st{s}$; if we make the ansatz that the sample conductivity relaxes with an exponential time dependence, then it follows that $1/\tau\st{fast}$ (and not $\tau\st{fast}$) should relax with an exponential time dependence.
We therefore assume that
\begin{subequations} 
\label{eq:taufast-model}
\begin{align}
\tau\st{fast}(t) 
   & = 1/k(t) \text{ and} \\
k(t)
  & = k\st{i} + (k\st{f} - k\st{i})(1 
   - e^{(t - t_0)/\tau_{\Gamma\st{s}}}),
\end{align}
\end{subequations}
with $t_0 = \SI{0}{\s}$ the time when the light was turned off, $k\st{i} = \tau\st{fast}^{-1}(t_0)$ an initial rate, $k\st{f} = \tau\st{fast}^{-1}(\infty)$ a final rate, and $\tau_{\Gamma\st{s}}$ the time constant describing the exponential relaxation of the conductivity in the dark.
The $t > t_0$ data of Figure~\ref*{dissipation-tip-pulse} was fit to Equation~\ref{eq:Gamma-model} and \ref{eq:taufast-model} with $k\st{i}$, $k\st{f}$ and $\tau_{\Gamma\st{s}}$ fit parameters.
The maximum dissipation $\Gamma\st{max}$ was obtained manually from the measured data and was not included as a fit parameter.

In the variable-temperature experiments described in the manuscript, a temperature diode mounted near the cantilever was used to monitor the temperature \cite{Smieska2015auga}.
The temperature was varied by placing the metal exterior of the probe in a cold bath or hot bath.
The quality factor $Q_{0}$ was measured at each temperature before illumination.
Where indicated, for clarity, $\Gamma\st{s}$ was normalized by dividing by the maximum value of sample-induced dissipation $\Gamma\st{max}$.

\section{Spatial variation in the BLDS response}

An attempt was made to map out the spatial variation of the BLDS response for the \ce{TiO2}-substrate sample.
In its current implementation, BLDS is a slow technique; acquiring \textit{each} data point in the curves of Figure~\ref{fig:BLDS-roll-off} took approximately $\SI{2}{\sec}$.
Ideally, one would like to record the sample's dielectric response at several frequencies at each location with the same spatial resolution as the associated atomic-force microscope (AFM) topography data.
To map out BLDS spectra across a $32 \times 30$ grid of sample locations, we reduced the signal-measuring time to $\approx$ $\SI{1}{\sec}$ per point.
BLDS spectra were acquired under constant illumination, $I_{h\nu} = \SI{1.5}{\milli\W\per\cm\squared}$, with the illumination intensity $I_{h\nu}$ chosen such that the response rolled off in the applied voltage-modulation frequency range. 
At each sample location, dielectric-response signal was acquired at $8$ voltage-modulation frequencies; this number of frequencies was chosen as a compromise, to minimize the time required to perform the measurement while still obtaining enough data points to infer the roll-off frequency.
At each location on the grid, AFM was performed and the tip was retracted by $\approx$ $\SI{50}{\nano\m}$.
In this way, a nominally constant tip-sample separation was maintained during the scan.
Figure~\ref{fig:BLDS-map}a is an AFM topography image acquired in the dark (after the BLDS measurement) which shows that the scanned area included a grain boundary.
In Figure~\ref{fig:BLDS-map}b, we plot the mean of $\alpha$ of the BLDS curve at each location on the grid.
The mean value of $\alpha$ is one way to visualize differences in the BLDS response.
This value is effected not only by the inherent differences in the dielectric response at that location but also by the tip-sample separation and the film thickness.
While every effort was made to keep the tip-sample separation constant, we can not rule out that the observed differences in $\alpha$ are due to point-to-point variations in film thickness.
To reveal whether the observed differences in $\alpha$ were due to a spatially-dependent roll-off frequency, in Figure~\ref{fig:BLDS-map}c we plot the full normalized dielectric response curves for $6$ points, separated by $\SI{150}{\nano\m}$, along the dotted line marked in Figure~\ref{fig:BLDS-map}b.
We see that the normalized dielectric response curves show \textit{qualitatively} similar frequency-dependent roll-off at different locations in the sample.
The apparent homogeneity in the dielectric response across a grain boundary could potentially be attributed to the tip-sample separation ($\approx$ $\SI{50}{\nano\meter}$) being comparable to the grain size ($\approx$ $\SI{200}{\nano\meter}$).
Samples with larger grain of more uniform film thickness will be better candidates to map out spatial variation in the BLDS response.

\begin{figure*}
\includegraphics[width=5in]{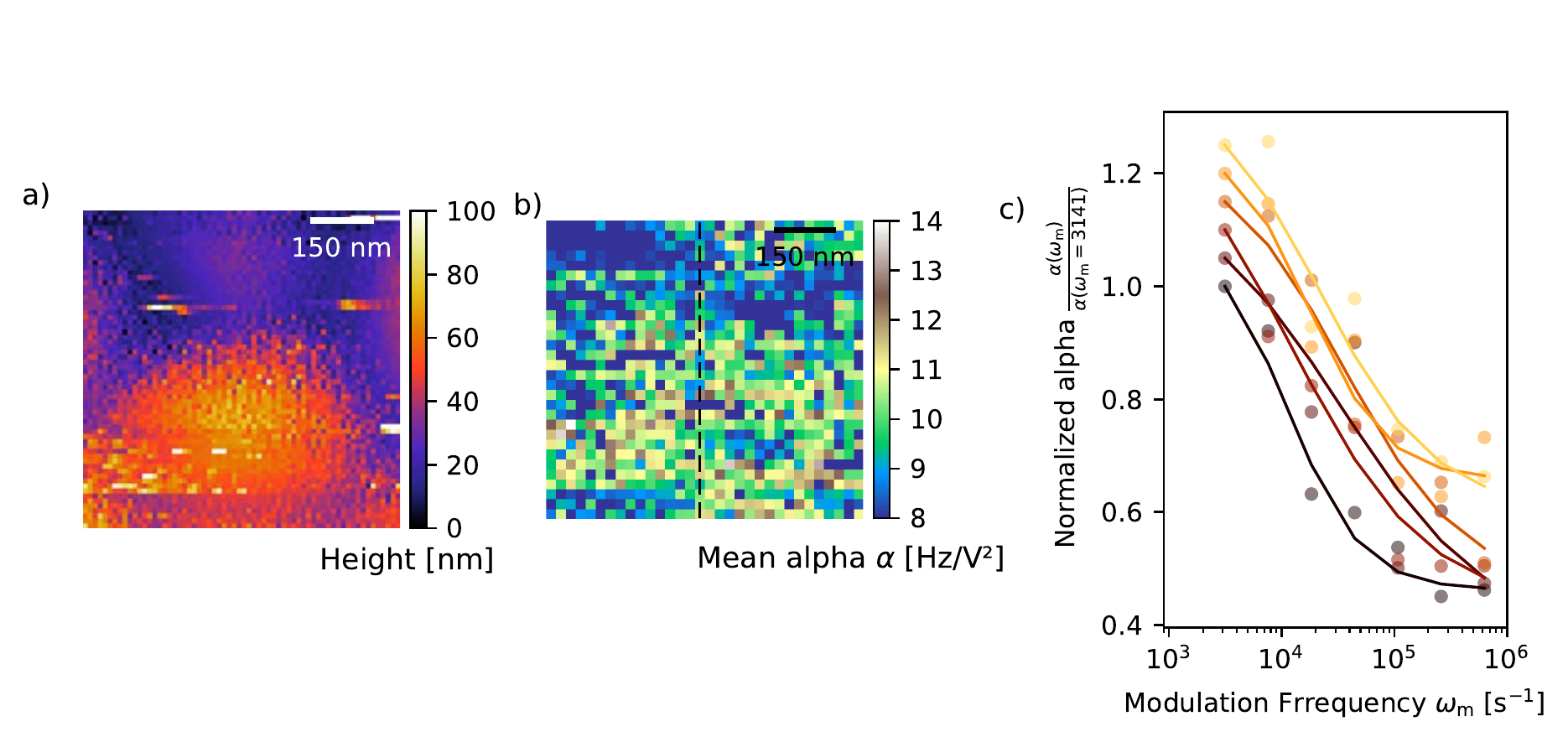}
\caption[Spatial dependence of the BLDS response (\ce{TiO2}-substrate)]{The BLDS response of the \ce{TiO2}-substrate sample is spatially homogeneous. 
(a) AFM topography image, showing that the mapped area contained a grain boundary. 
(b) Mean alpha $\alpha$ calculated from the BLDS curve acquired at each location in the $32 \times 30$ tip-scan grid.
(c) Normalized BLDS curves (each offset vertically by $0.05$) acquired at $6$ locations along the dotted line in (b); the point separation is $\SI{150}{\nano\meter}$.
Solid lines are a fit to a one-time-constant low-pass filter; these lines are presented only as a guide to the eye.
In (c), each data set is colored according the cantilever-tip location along the dotted line in (b), going from bottom (dark) to top (light).
}
\label{fig:BLDS-map}
\end{figure*}

\section{Supplementary figures summary}
Figure~\ref{fig:SI-Amplitude-Freq-example} shows example frequency shift ($\Delta f$) \latin{vs.}~applied tip voltage ($V\st{ts}$) and (b) amplitude ($\Delta A$) vs \latin{vs.}~applied tip voltage ($V\st{ts}$) data for the data plotted in Figure~\ref{fig:Dissipation-curvature} at selected light intensities for the \ce{TiO2}-substrate sample.
Figure~\ref{AFM-combined} shows representative AFM images of the four films studied in the manuscript.
Figures~\ref{surface-potential} and \ref{transient-surface-potential} show the surface potential of each film \latin{vs}.\ light intensity and time, respectively.
We note that the surface potential in each of these samples showed a distinct, non-trivial light-dependent and time-dependent behavior. 
Figure~\ref{Dissipation-below-bandgap} shows that if below-bandgap illumination was used, there was no change in sample-induced dissipation, confirming that generation of free carriers is necessary to observe a light-induced change in sample conductivity.
Figure~\ref{SnO-high-temp} shows that the dissipation-recovery time constant $\tau_{\Gamma\st{s}}$ for the \ce{SnO2}-substrate sample did not change in the measured temperature window.
\begin{figure*}
\includegraphics[width=3.0in]{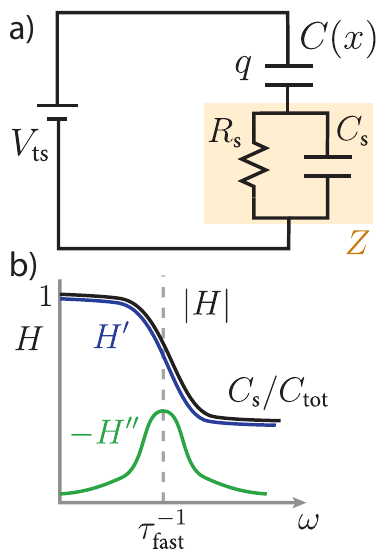}
\caption[Equivalent circuit and transfer function]{The equivalent circuit for the cantilever and sample and the associated response function; absolute value $H$ (upper, black) real part, $H'$ (upper,blue), and imaginary part $H''$ (lower, green). Adapted with permission from Ref.~\citenum{Tirmzi2017jana}. Copyright (2017) American Chemical Society.} 
\label{fig:SI-equivalent-circuit-old-paper}
\end{figure*}

\begin{figure*}
\includegraphics[width=5in]{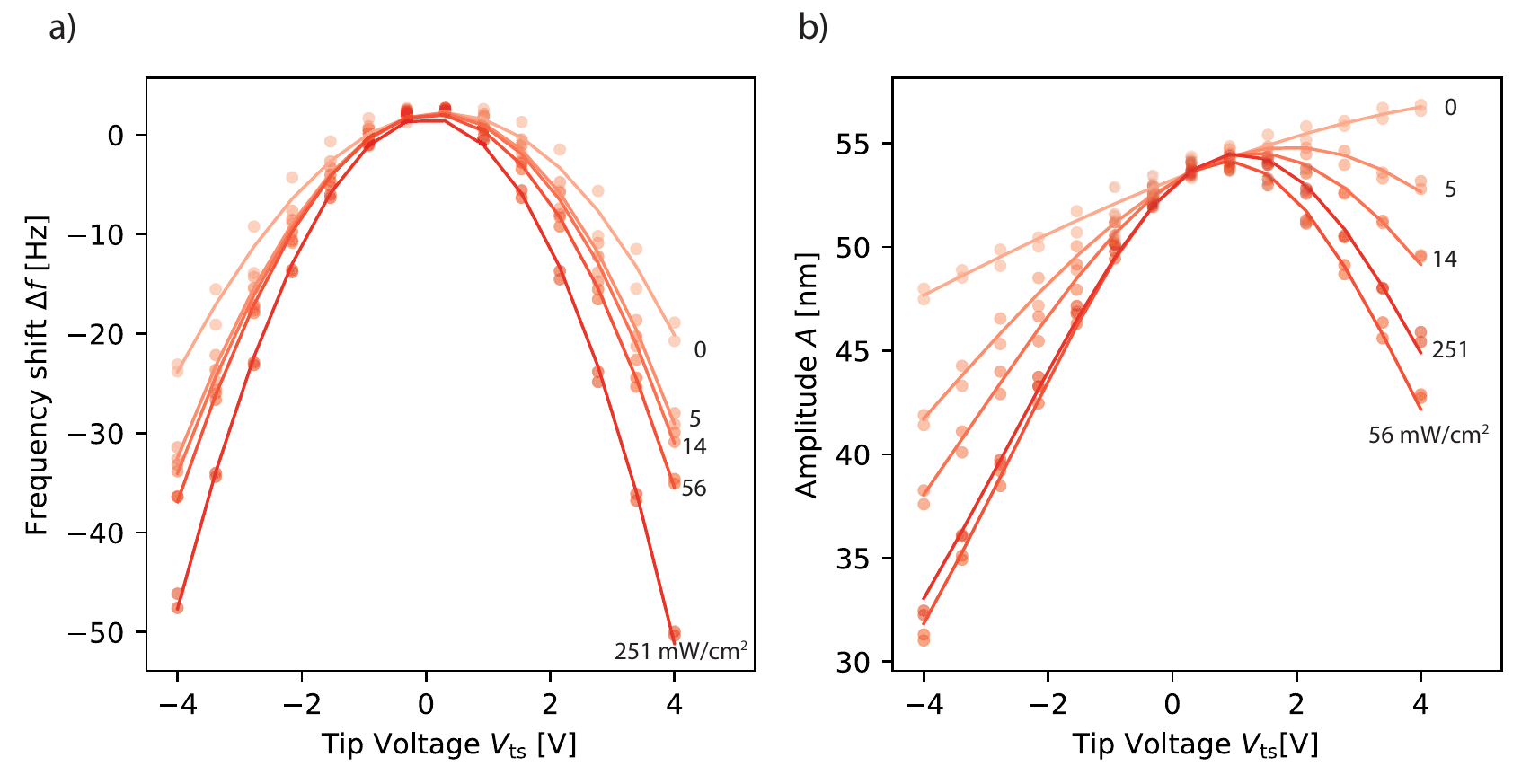}
\caption[Example data for the Figure~\ref{fig:Dissipation-curvature}]{Representative (a) frequency shift ($\Delta f$) \latin{vs.}~applied tip voltage ($V\st{ts}$) and (b) amplitude ($\Delta A$) \latin{vs.}~applied tip voltage ($V\st{ts}$) data for Figure~\ref{fig:Dissipation-curvature} at selected light intensities. Curvature ($\alpha\st{0}$) is calculated from a fit to Eq.~\ref{eq:fq-vs-voltage} of the frequency shift data in (a). Voltage-normalized sample-induced dissipation ($\gamma\st{s}$)is calculated from a fit to Eq.~\ref{eq:gamma-s-vs-voltage} of the amplitude data shown in (b). See Section~\ref{sec:curvature-dissipation-calculation} for details regarding the calculations.} 
\label{fig:SI-Amplitude-Freq-example}
\end{figure*}
\begin{figure*}
\includegraphics[width=5in]{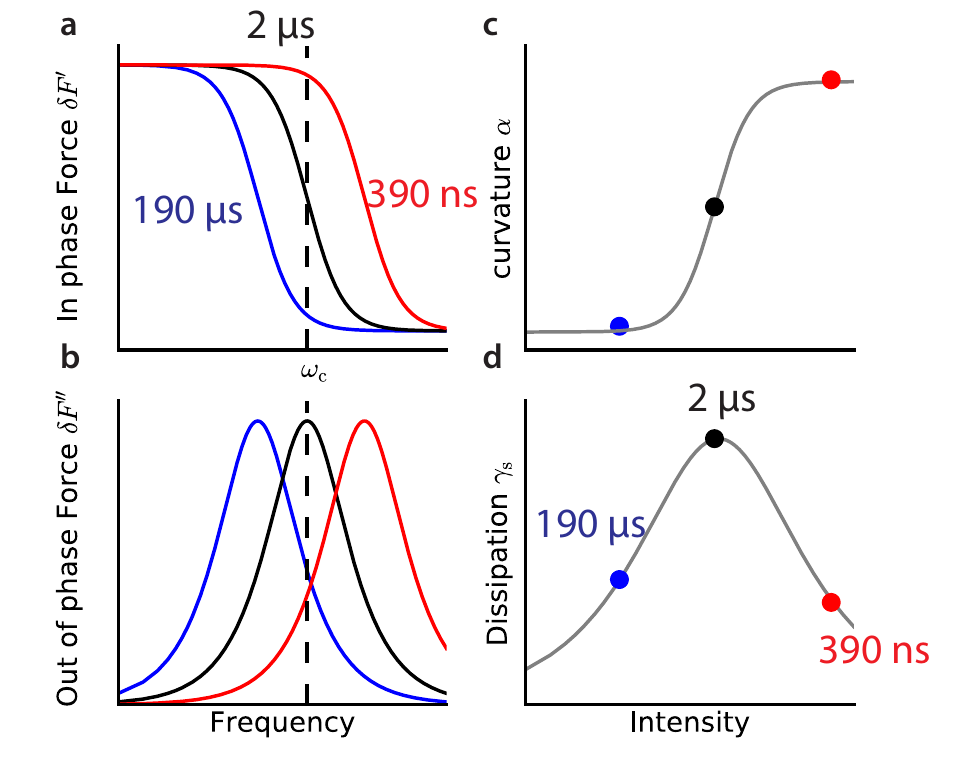}
\caption[Predicted behavior for curvature and dissipation with light]{Predicted behavior for curvature and dissipation with light. (a) In-phase and (b) out-of-phase force for various values of $\tau\st{fast}$ according to the transfer function described by Eq.~\ref{eq:H}: (left, blue curve) slow $\tau\st{fast}$, low $I\st{h\nu}$; (middle, black curve)
intermediate $\tau\st{fast}$ , intermediate $I\st{h\nu}$; (right, red curve) short $\tau\st{fast}$, $I\st{h\nu}$. A peak in dissipation is seen when $\tau\st{fast}$ $= \SI{2}{\micro\s}$. Figure~\ref{fig:Dissipation-curvature} measurements probe the in-phase and out-of-phase forces at the cantilever frequency, $\Delta F'$($\omega\st{c}$) and $\Delta F''$($\omega\st{c}$) ; the cantilever frequency is indicated as a dotted vertical line in panels (a) and (b). The predicted intensity dependence of the (c) frequency-voltage curvature ($\alpha\st{0}$) and (d) voltage-normalized sample-induced dissipation ($\gamma\st{s})$) agrees
with the Figure~\ref{fig:Dissipation-curvature} data. Adapted with permission from Ref.~\citenum{Tirmzi2017jana}. Copyright (2017) American Chemical Society.} 
\label{fig:cartoon-old-paper}
\end{figure*}
\begin{figure*}
\includegraphics[width=5.0in]{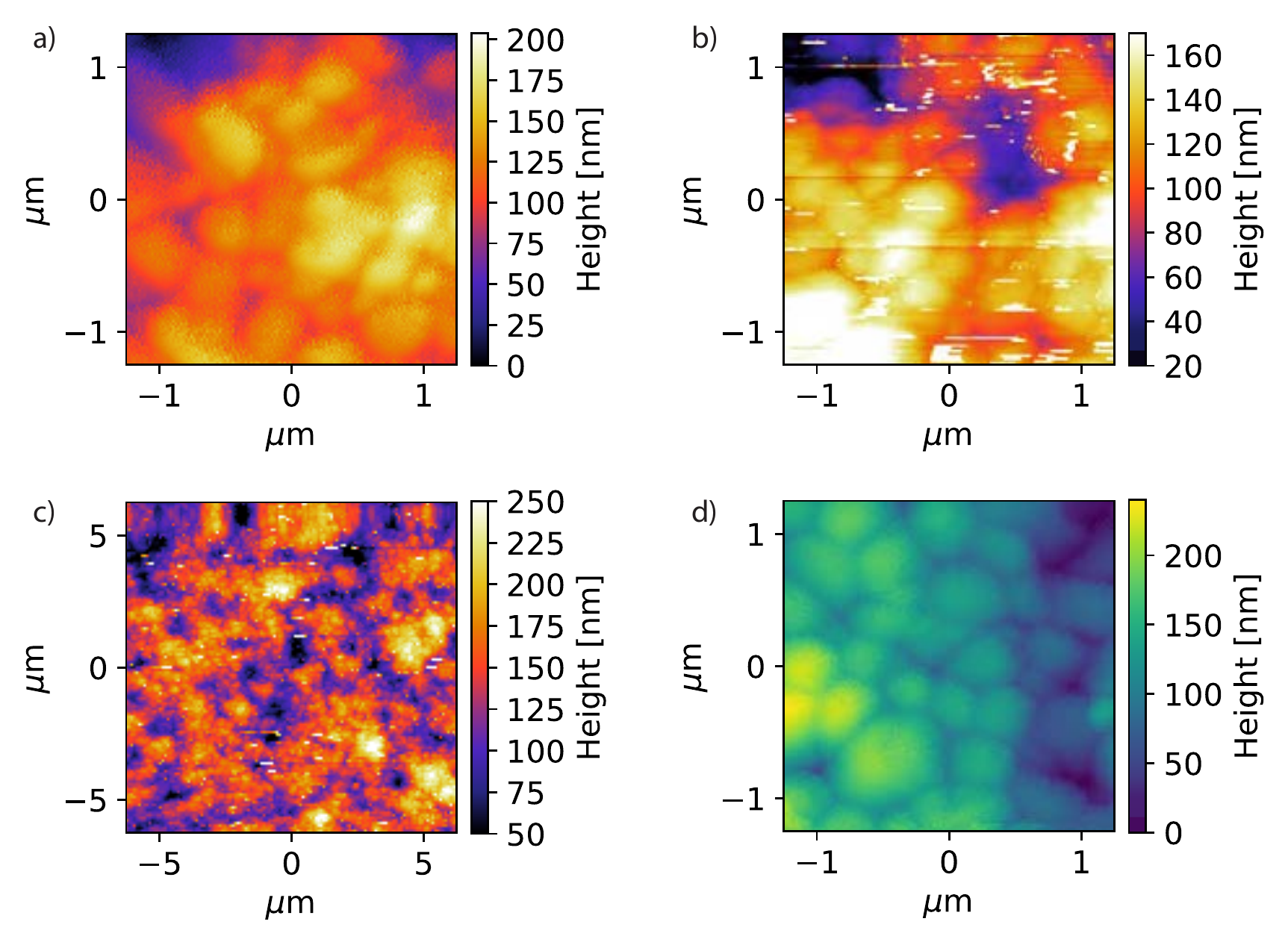}
\caption[AFM topography (four substrates)]{Representative AFM topography images for FAMACs films prepared on (a) \ce{TiO2}, (b) \ce{SnO2}, (c) \ce{ITO}, and (d) \ce{NiO}.}
\label{AFM-combined}
\end{figure*}

\begin{figure*}
\includegraphics[width=3.5in]{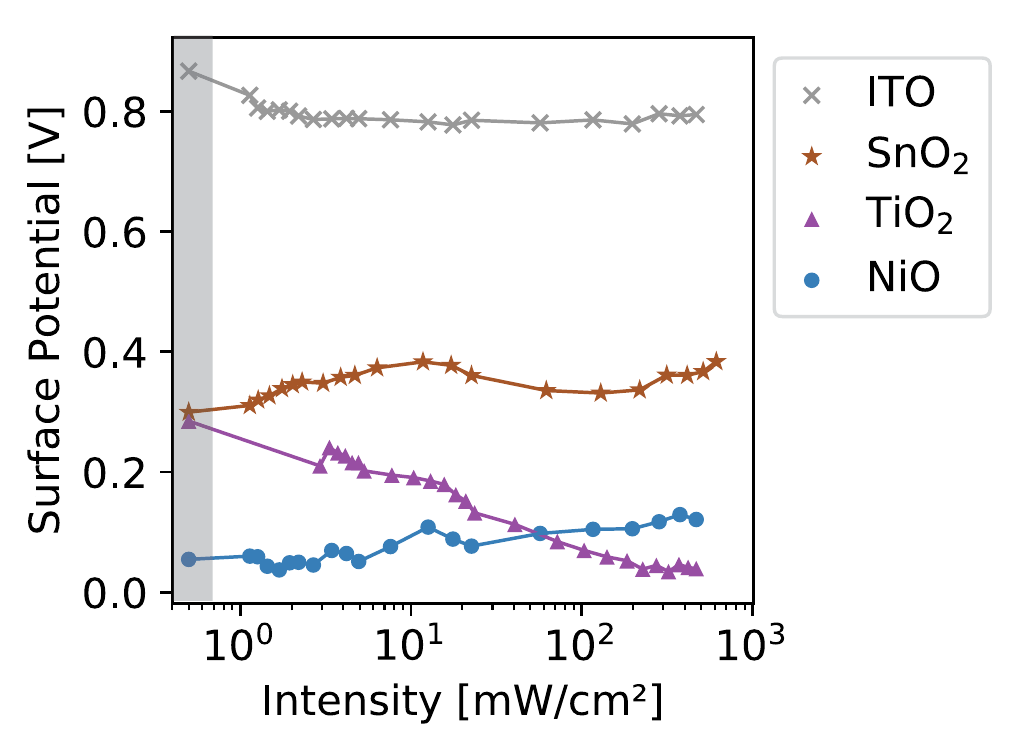}
\caption[Surface potential \latin{vs}.~light intensity (four substrates)]{Quasi-steady-state surface potential under illumination for FAMACs thin films prepared on the indicated substrates. 
	The leftmost, shaded data point in each surface-potential curve corresponds to the surface potential in the dark.
	For experimental parameters please refer to Figure~\ref{fig:Dissipation-curvature}.}
\label{surface-potential}
\end{figure*}

\begin{figure*}
\includegraphics[width=5in]{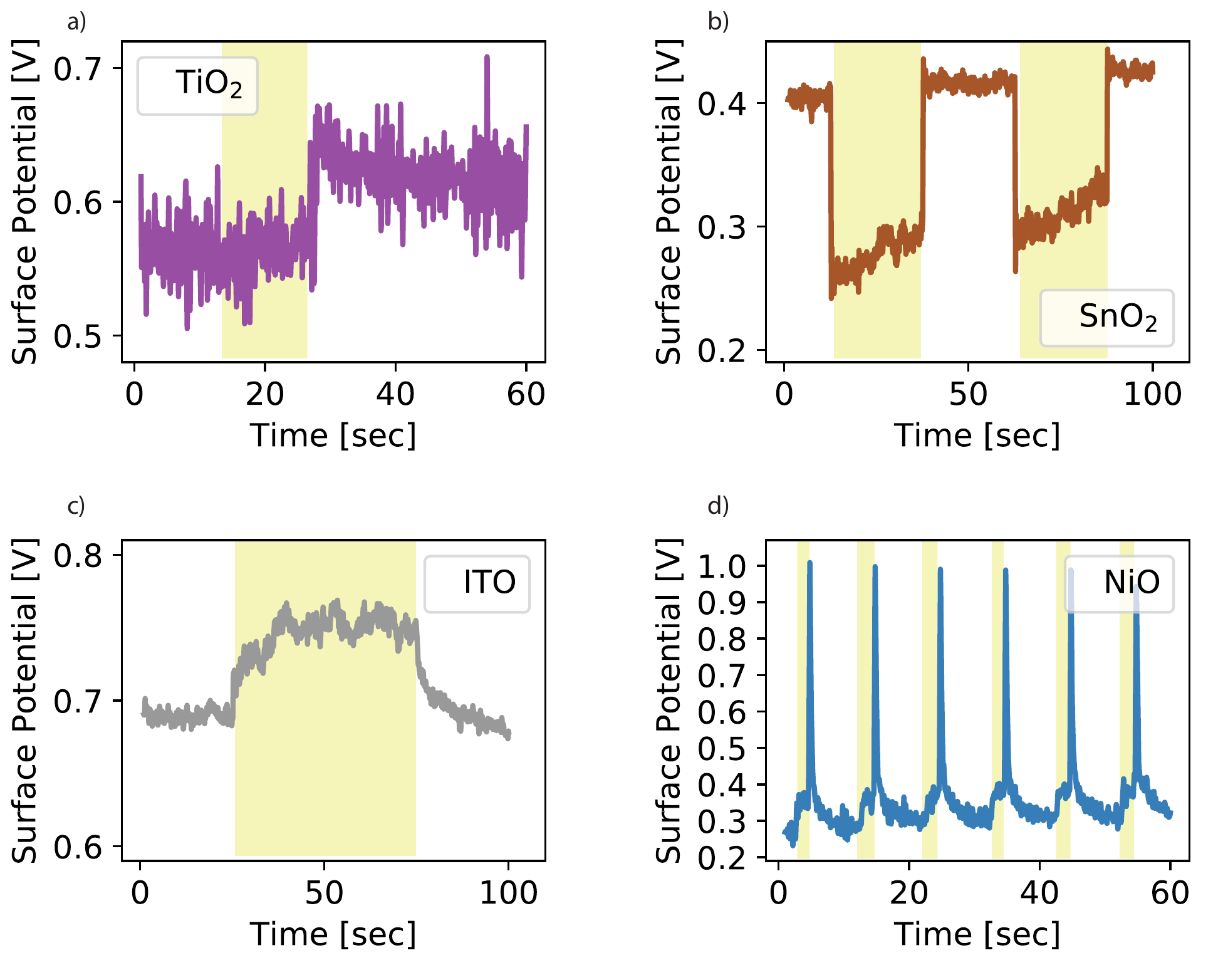}
\caption[Surface potential \latin{vs.}~time (four substrates)]{Transient surface potential of FAMACs films prepared on the four substrates.
	The illumination period is represented by a yellow shaded region.
	Sample identity and illumination intensity: (a) \ce{TiO2}, $I\st{h\nu} =\SI{5}{\milli\W/\cm\squared}$;  
(b) \ce{SnO2}, $I\st{h\nu} =\SI{321}{\milli\W/\cm\squared}$; 
(c) ITO, $I\st{h\nu} =\SI{321}{\milli\W/\cm\squared}$; and 
(d) \ce{NiO}, $I\st{h\nu} =\SI{5}{\milli\W/\cm\squared}$.}
\label{transient-surface-potential}
\end{figure*}

\begin{figure*}
\includegraphics[width=3.5in]{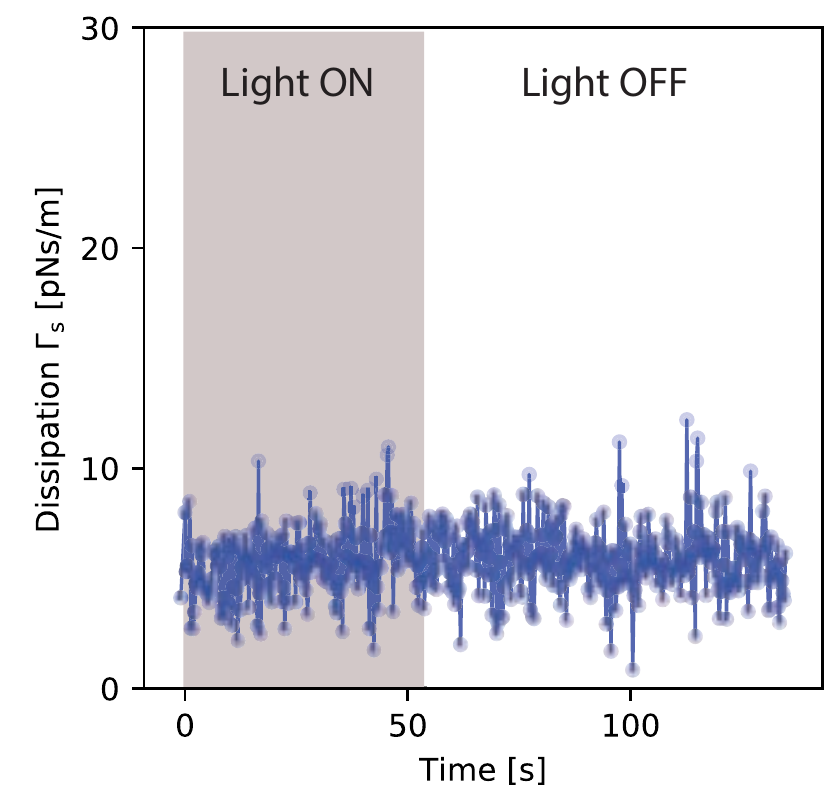}
\caption[Change in dissipation using below-bandgap illumination (\ce{TiO2}-substrate)]{No light-induced sample dissipation was observed when the \ce{TiO2}-substrate sample was illuminated with below band-gap excitation.
	Light was turned on at time $t = \SI{0}{\sec}$ and was turned off at time $t = \SI{54}{\sec}$.
	Experimental parameters: $\lambda =$ $\SI{980}{n\m}$, $V\st{ts} = \SI{-4}{\V}$, $T = \SI{292}{\K}$, $h = \SI{150}{nm}$, and $I_{h\nu} = \SI{300}{\milli\W\per\cm\squared}$.}
\label{Dissipation-below-bandgap}
\end{figure*}

\begin{figure*}
\includegraphics[width=5.0in]{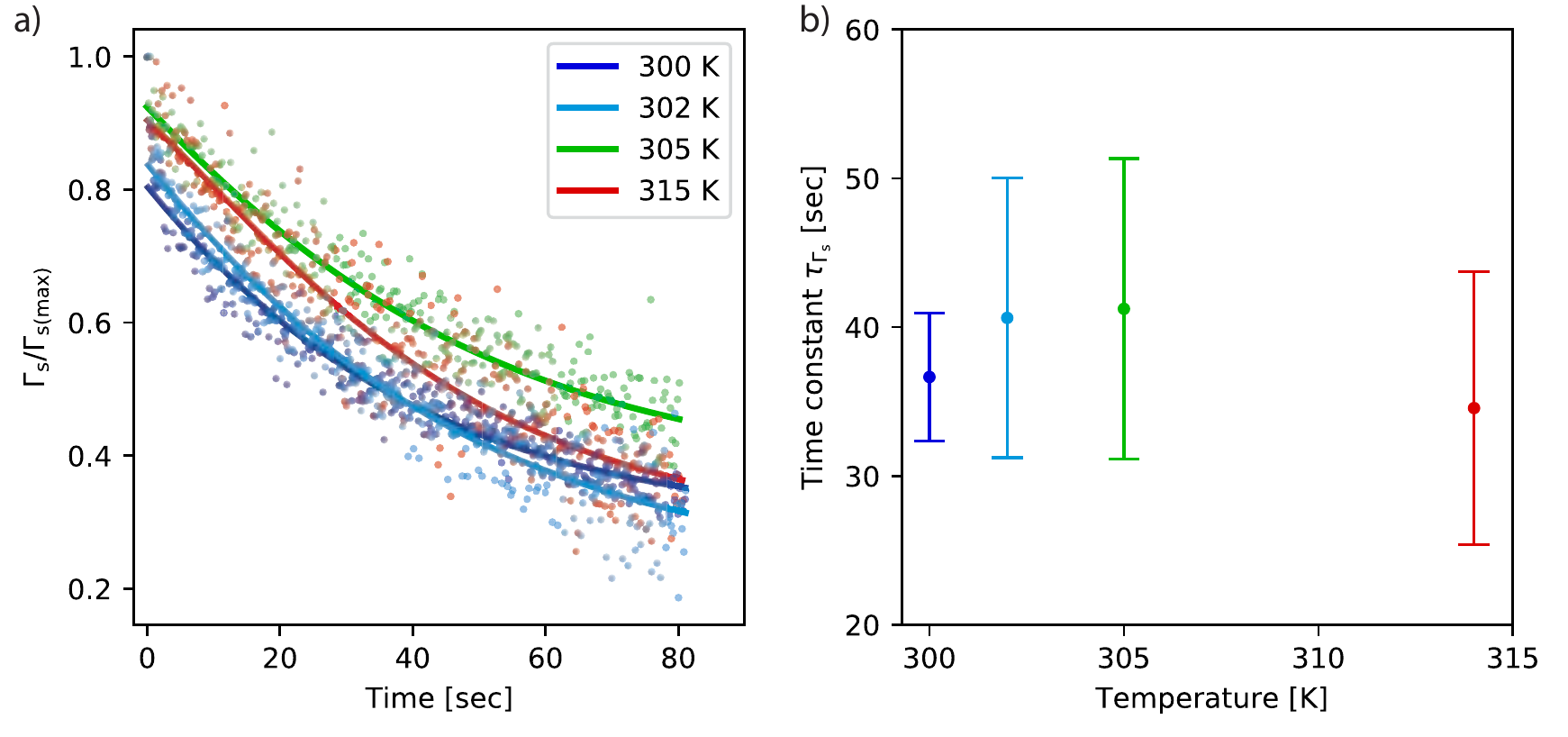}
\caption[Dissipation-recovery time constant \latin{vs.}~temperature (\ce{SnO2}-substrate)]{The dissipation-recovery time constant for the \ce{SnO2}-substrate sample in the $\SI{300}{\K}$ to $\SI{315}{\K}$ temperature range.
	(a) Dissipation-recovery transient and best fits to a single-exponential model. 	(b) Time constant obtained from the exponential model ($\pm1 \: \sigma$ error bars).
	We attribute the large experimental error to an increased sample-induced dissipation at elevated temperature in this sample.
	The light intensity was $I\st{h\nu} =\SI{1.5}{\milli\W/\cm\squared}$. }
\label{SnO-high-temp}
\end{figure*}

\begin{figure*}
\includegraphics[width=5in]{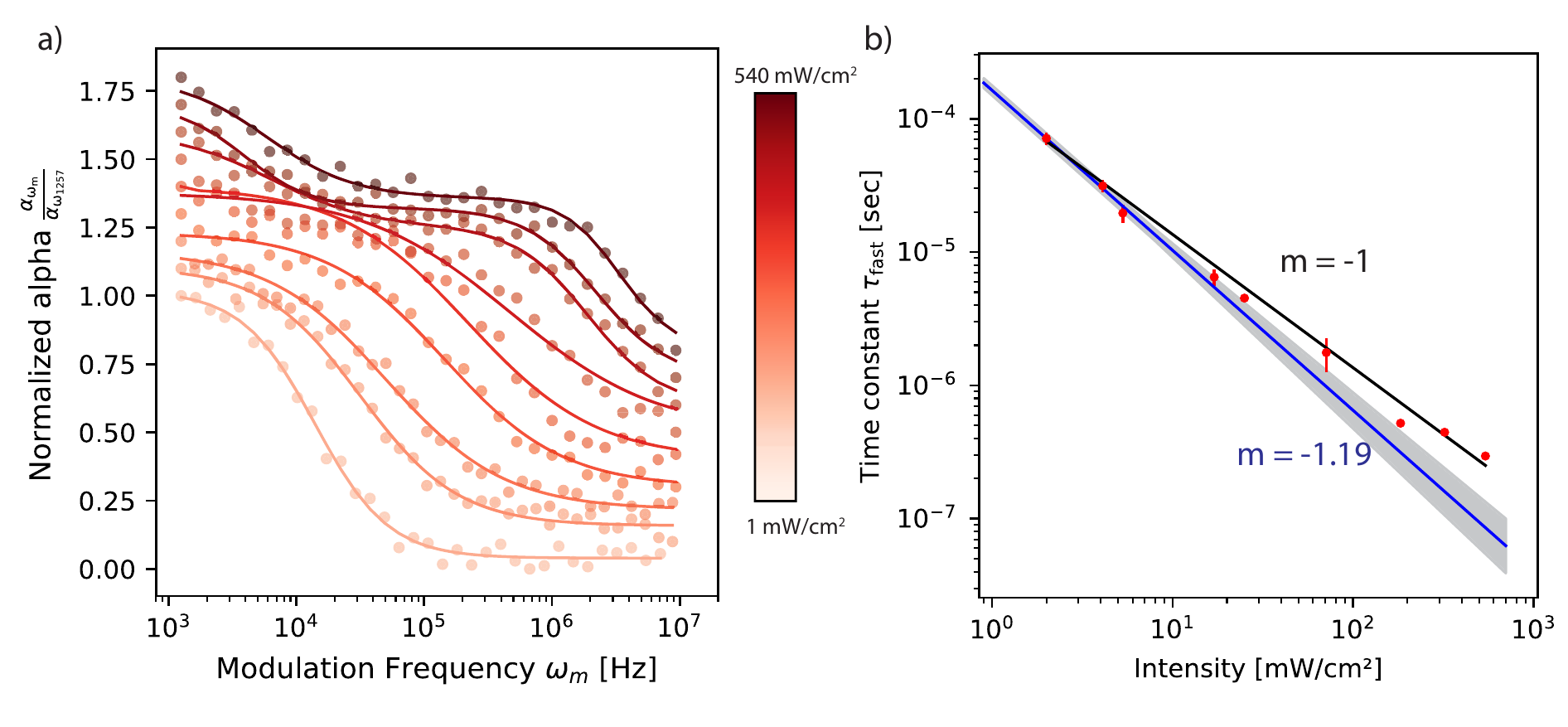}
\caption[Time constant calculated from BLDS for \ce{TiO2}-substrate sample]{Fast time constant calculated from a fit to a single time constant or a two time constant low pass filter of BLDS data for \ce{TiO2}-substrate sample.
	(a) BLDS curves for \ce{TiO2} substrate sample normalized to the first data point and vertically off set by 0.1 (same experimental data as Figure~\ref{fig:BLDS-roll-off}). 	(b) Fast time constant ($\tau\st{fast}$) obtained from the fits in (a) is plotted \latin{vs.}~light intensity. Blue solid line shows fit to $y = b \times \{\frac{I\st{h\nu}}{I\st{0}}\}^{m}$ where $m = -1.19$ and $I\st{0}$ is fixed $\SI{1}{\milli\W\per\cm\squared}$. Black solid line is guide to the eye for $m = -1$ fit.}
\label{TiO-time-constant}
\end{figure*}

\begin{figure*}
\includegraphics[width=5in]{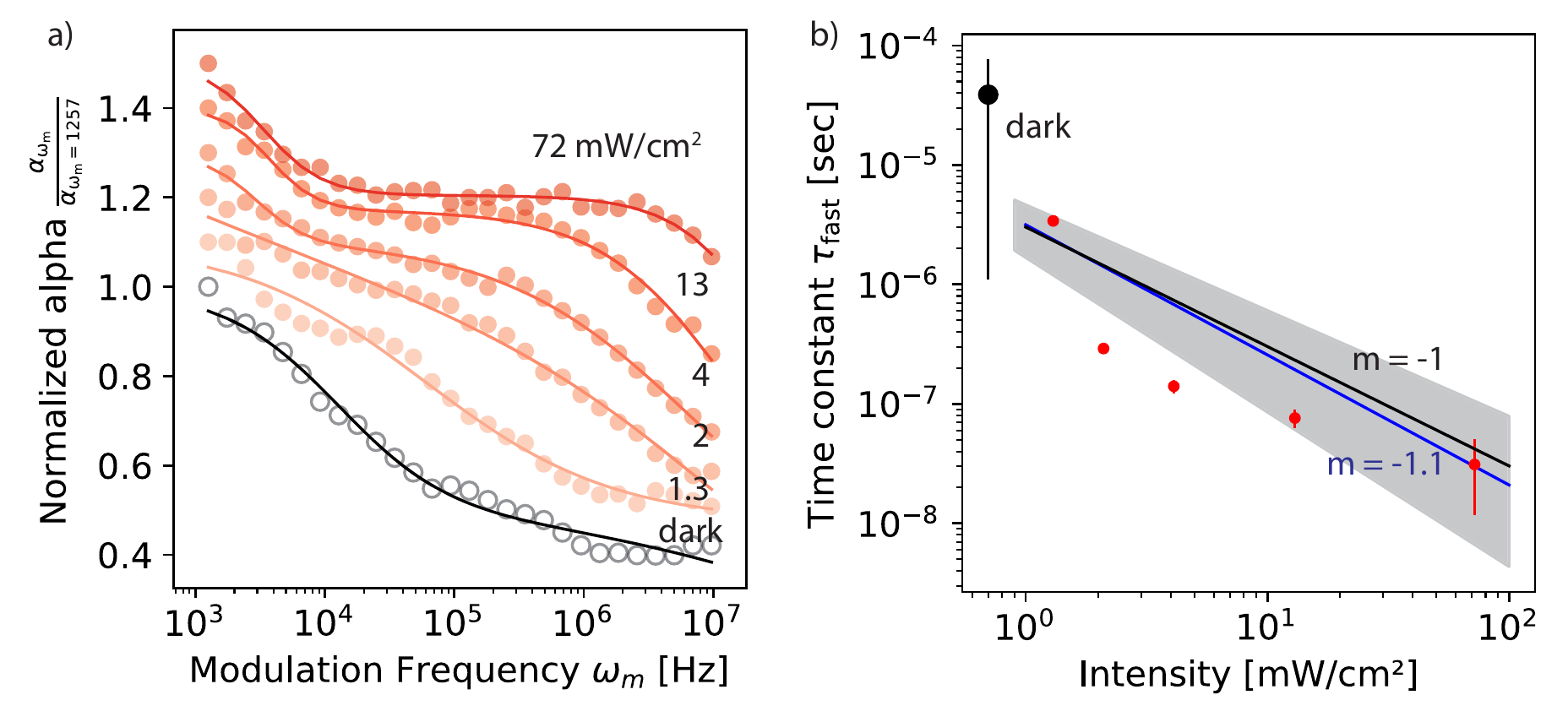}
\caption[Time constant calculated from BLDS for \ce{SnO2}-substrate sample]{Fast time constant calculated from a fit to a two time constant low pass filter of BLDS data for \ce{SnO2}-substrate sample.
	(a) BLDS curves for \ce{SnO2} substrate sample normalized to the first data point and vertically off set by 0.1 (same experimental data as Figure~\ref{fig:BLDS-roll-off}). 	(b) Fast time constant ($\tau\st{fast}$) obtained from the fits in (a) is plotted \latin{vs.}~light intensity. Blue solid line shows fit to $y = b \times \{\frac{I\st{h\nu}}{I\st{0}}\}^{m}$ where $m = -1.1$  and $I\st{0}$ is fixed $\SI{1}{\milli\W\per\cm\squared}$. Black solid line is guide to the eye for $m = -1$ fit.}
\label{SnO-time-constant}
\end{figure*}

\end{suppinfo}

\label{TheEnd}

\end{document}